\documentclass{ws-mpla}
\usepackage{latexsym,color}
\usepackage{hyperref}
\begin{document}
\newcommand{\ti}[1]{\mbox{\tiny{#1}}}
\newcommand{\im}{\mathop{\mathrm{Im}}}
\def\be{\begin{equation}}
\def\ee{\end{equation}}
\def\bea{\begin{eqnarray}}
\def\eea{\end{eqnarray}}
\newcommand{\tb}[1]{\textbf{\texttt{#1}}}
\newcommand{\rtb}[1]{\textcolor[rgb]{1.00,0.00,0.00}{\tb{#1}}}
\newcommand{\ptb}[1]{\textcolor[rgb]{0.65,0.29,0.71}{\tb{#1}}}
\newcommand{\il}{~}
\newcommand{\Tem}{T^{\ti{(em)}}}
\newcommand{\Tm}{T^{\ti{M}}}
\newcommand{\apis}[1]{\ ^{\mbox{\tiny{(#1)}}}\! }%
\newcommand{\Tkg}{T^{\ti{(KG)}}}
\newcommand{\g}[1]{\Gamma^{\phantom\ #1}}
%
%

\markboth{D. Pugliese, G. Montani}
{Astrophysical evidence for an extra dimension: phenomenology of a Kaluza Klein theory}

\catchline{}{}{}{}{}

\title{Astrophysical evidence for an extra dimension: phenomenology of a Kaluza Klein theory
}

\author{\footnotesize D. Pugliese}

\address{School of Mathematical Sciences, Queen Mary University of London \\
 Mile
End Road, London E1 4NS, UK, EU.\\
dpugliese@maths.qmul.ac.uk}

\author{G. Montani}

\address{ ENEA - C.R, UTFUS-MAG -
Via Enrico Fermi 45, 00044
Frascati, Roma, Italy, EU\\
Dipartimento di Fisica, Universit\`a di Roma ``Sapienza"\\
 Piazzale Aldo Moro 5, I-00185 Roma, Italy, EU\\
giovanni.montani@frascati.enea.it
}

\maketitle

\pub{Received (Day Month Year)}{Revised (Day Month Year)}

\begin{abstract}
In this brief review we discuss the viability of a multidimensional geometrical theory with one compactified dimension.
We discuss the case of a  Kaluza Klein fifth dimensional theory, addressing the problem by  an overview of the astrophysical phenomenology   associated with this  five dimensional theory. By comparing   the predictions of our model with the features of  the  ordinary (four dimensional) Relativistic Astrophysics, we highlight some small but finite discrepancies,  expectably detectible from the observations.
We consider a class of static, vacuum  solutions of free electromagnetic
Kaluza Klein equations with three dimensional spherical symmetry. We explore the stability of the particle dynamics in these spacetimes, the construction of self gravitating stellar models  and the emission spectrum generated by a charged particle falling on this stellar object.
The matter dynamics  in these geometries  has been treated by a multipole approach adapted to the geometric  theory with a compactified dimension.
\end{abstract}

\ccode{PACS Nos.:04.50.Cd, 04.50.Gh.}

\section{Introduction}
Geometrical extra dimensional  theories represent a good
candidate for the GUT (Great Unification Theory)
\cite{Bergamini:1984gx,Aranda:2009wh}, being often used in this framework in  attempt to find a unified theory of the four fundamental forces  through the geometrization of the other fields in addition to the gravity as described by General Relativity. These theories are also used  in the development of theoretical models of the evolution of our Universe as emerging from the observation, in fact many    cosmological theories  have been derived, considering  an implementation of an higher  dimension number   than the four  of the original Einstein model (\cite{Gunther:2002mm,Brax:2003fv,Langlois:2002bb,Papantonopoulos:2002ew,Gun.Star.Zhuk04}).

On the other hand, the recent results of the LHC experiments are still the subject of many debates and theoretical studies aimed at finding a data interpretation in  terms of a possible  multidimensional scenario compatible with the recent constraints (see for example \cite{Ar-Gi-Smo-Yo-2013,Sun:2012xt,Choudhury:2011jk,deBlas:2012qf,Zhou-Fei-Gan-You-Lei-2012,Gerwick:2011jw,Arrenberg:2008wy,Calmet:2009yw}).
Therefore  there is a great interest  in providing a theoretical model able to explain the role of the
extra dimensions and their meaning  in a world that looks like a
four dimensional one, but up to now no experimental evidence could outline a signature   of extra dimensions but they could simply fix restriction on  the existing proposals (see for example \cite{Matsuno:2011ca,Rizzo:1999qb,Eingorn:2010wi,Moutsopoulos:2011ez,Stelea:2009ur,Bonnevier:2011km,Okawa:2011fv,Tomizawa:2011mc}
 and also \cite{Moon:2011sz,Becar:2011fc,Yamada:2011br,Inte}).

This paper  participates   this discussion by presenting a series of recent developments on the possible observational scenarios  provided as available from  the Universe observation, that could lead to an analysis of the observed phenomena in terms of multi dimensional models.

In \cite{Lacquaniti:2010zz,Lacquaniti:2009rh,Pugliese:2011um,Pugliese:2011yh,Lacquaniti:2009wc}  it has been  taken into account  the dynamics of moving bodies subjected to strong gravitational fields,  and it was studied  the test particle motion in view of further developments in the exploration of the   extended matter behavior in this  context, as the accretion disks around compact objects in a possible multi dimensional Universe.
Actually we explore the very distant objects  by the observation of their emission spectra, and the jets in the electromagnetic band like x ray or gamma ray etc. Many of these phenomena, like GRBs   for example, are today still being explored. The  origin  on these high  energy phenomena is generally associated with the initial stages of  the development in the birth of a black hole, but they are far from being well understood  in a  satisfactory  model.
We look at these phenomena as fruitful environment in which we can highlight the details  attributable in general to   theories alternative  to the  Einstein's one.
Finally as outlined in \cite{Pugliese:2011yh} the study of the multidimensional  stellar structure could be   an interesting investigation area  of  the main astrophysical implications
of the  higher dimensional gravity. This extra dimensional star model should provide  a usually star as  observed in the apparent   four dimensional (4D) Universe, but with some peculiar characteristics  due to the presence of an extra dimension which regulates the structure, the equilibrium configuration and possibly its  evolutionary history  towards its final stages, see also \cite{PaulBC,Hannestad:2003yd,Patel:2001jw,cha}.

However besides  the problem of  the experimental proof of the existence of a possible extra dimension there are several subtle issues related to these theories.
Firstly it should be clearly explained  why one does not perceive the  presence of one or more extra dimension as provided by the   theory.
There are several viable proposals  to this question, and the  two main different elaborations of this problem consists on  one hand in the assumption of a very large extra dimensional size  and  on the other hand the conjecture of a  very small extra dimension, that is  below the  Planck scale\cite{K1,K2,K3}, for  example this is the case of the  5D Kaluza Klein (KK) models.
This is accomplished in  different ways, however, we refer in this paper to the compactification of the extra dimension.
In general, the  Kaluza Klein models provide the geometrization of
the electromagnetic field  coupled with the gravitational one.
In this review we consider a free electromagnetic  5D KK model   with  a scalar field associated to the extra dimensional component of the metric  which regulates  its the size  and dynamics.
These theories are characterized by the  gauge invariance (spacetime symmetry)
implemented by the invariance for
translations on the compactified fifth dimension, representing   the internal U(1) gauge symmetry via an isometric group.
 Nevertheless,  the  test particle dynamics constitutes a significant shortcoming
known as the charge to mass puzzle:  because  of  the huge massive modes arising  near the Planck scale, it is possible to recover
 the  particle motion but at the cost of the charge to mass
ratio  not matching with any observed particle.

It was recently proposed a revision of the dynamics in KK theories with compactified dimensions
that seems to solve this problem without affecting the correct representation of the reduced 4D physical  Universe
\cite{Lacquaniti:2009yy}: specifically it has been  taken into account a  definition of a 5D particle as a
localized matter distribution in the ordinary 4D spacetime but as a
delocalized one  on the fifth dimension. This starting point   leads to a
different definition of mass which solves the charge to mass puzzle: the procedure of the dimensional reduction from the  5D to 4D dynamics is  based on a Papapetrou multipole expansion
of a 5D energy momentum tensor which is supposed to be picked along
a 4D world tube.
Here  we  take into account   a family of  solutions of the KK equations  in vacuum, known as Generalized
Schwarzschild spacetime
(GSS)\cite{Lacquaniti:2010zz,Overduin:1998pn,Wesson:1999nq}.
These metrics are generally interpreted as an extension of the Schwarzschild spacetime  to a 5D scenario,  because in appropriate limits on the family parameters  on the slice $x_5 =\rm{const}$ (the 4D metric counterpart), the GSS tensor reduces  to the Schwarzschild metric.

The particle motion  has been addressed in \cite{Lacquaniti:2010zz} using an effective potential approach. The test particle dynamics, circularly  orbiting  around a compact object  is studied as a  1D nonrelativistic motion,  governed by an effective   potential which encloses the gravitational field, the centrifugal force and the effects due to the size  variation of the extra dimension.
We  review  the morphology of the circular motion by  fixing  the last circular orbit radius and  the last stable circular orbits radius of  charged and
neutral test particles.
These studies show that even in this simple spacetime model, some small but finite deformations  of the physics with respect to the prediction of the   4D model  appear.
It is finally interesting to note that a  deformation was found in \cite{Pugliese:2011um} in the emission spectrum generated by a charged particle falling  in a GSS towards the surface of a stellar object.
Therefore the extra dimensional phenomenology of an Astrophysical setting  could be an valuable environment in which  searching for the signature  of an extra dimension presence.
In fact, the additional extra dimensional
produces a nontrivial departure from the dynamics in the
corresponding 4D counterpart of the GSS. We believe that this modification from the results expected in the 4D model can be even more evident in the dynamics of the extended matter configurations, for example  in accretion disk scenarios.

The aim of this review is to focus attention
on the possibility to {highlight} finite and observable
effects from the presence of an extra dimension,
by using an astrophysical scenario. The basic idea
is that the emitted radiation is produced
in a region of the space, near the compact 5D star,
where this dimension is not compactified on
a Planckian scale and therefore it affects the
phenomenology of the emitting source, but
there features are detected at very large distance,
where the space resembles an Euclidean (asymptotically
flat) one endowed with an unobservable
compactified dimension. In this sense, the
astrophysical setting offers the most natural arena
for investigating the phenomenology of extra dimensions:
the elementary particles (photons or more in general cosmic rays)
are detected in the ordinary four dimensional
Minkowski spacetime, where no direct measurement
of the additional dimensions could be physically
allowed, but they bring together information of the
multi dimensionality of the spacetime region where it
was produced. The present analysis, putting together
different results present in the literature and
providing them with a precise link, defines
the necessary algorithm to use this powerful investigation
scenario. The fundamental steps can be summarized as
follows: i) a satisfactory paradigm to interpret
the four dimensional matter (particle) phenomenology
in term of the geometrical structure of the extra space,
ii) the multidimensional version of a compact
astrophysical object (star or black hole);
associated with an interior or vacuum solution of the
spherically or axially symmetric problem;
iii) the detailed calculation of the particle
energy spectrum as predicted by the phenomenology
nearby the surface (or horizon) of such generalized sources.

Limiting the attention to a pure 5D Kaluza Klein
scenario, this review successes in achieving
all these three steps and therefore it constitutes
a valuable framework, following which more general
investigation can start up.

This paper is structured as follows: In Sec.\il(\ref{Sec:5dgenerico}) we illustrate the main characteristics of the 5D KK model   and  we discuss  the test particle motion in the  KK theory according to   the standard approach. In Sec.\il(\ref{Sec:papaint}) the  alternative scenario for the matter behavior in KK compactified theories  is introduced by  Papapetrou's approach. We discuss the  resulting formulation of the motion  and the  KK field equation.  Generalized Schwarzschild  spacetime (GSS)  is introduced  in Sec.\il(\ref{Sec:GSS}). Test particle dynamics in this spacetime is explored in Sec.\il(\ref{Sec:CircularorGSS}). Star models in the   5D KK theories are introduced in Sec.\il(\ref{Sec:KK-Stars}). The study of the spectral emission of charged particles radially infalling in a GSS spacetime is discussed  in Sec.\il(\ref{Sec:perturbation}). Finally  Sec.\il(\ref{Sec:concldisc}) closes this paper with the discussions and future perspectives.


\section{Five dimensional Kaluza Klein model}\label{Sec:5dgenerico}
In  this paper  we investigate  the astrophysical systems  in the scenario envisaged by the 5D KK models, and looking for evidence that the  4D Universe would emerge as a dimensional reduction of  the  KK spacetime.
Therefore  this KK hypothesis
requires  the process of the  dimensional reduction
be satisfactory addressed.
The fundamental requirement to make unobservable
the fifth dimension is the existence of a closed
topology
which allows to compactify the spacetime.
The basic request to deal with a closed
topology is the periodicity of the metric field on
the fifth coordinate, which allows to expand the
Einstein Hilbert action in  Fourier series and
so to develop the dynamics of the different harmonics.
This approach is intrinsically different from
the cylindricity condition, due to Kaluza,
which prescribes the independence of the metric tensor
by the fifth coordinate, directly within the field
equations. However,  if  we truncate the
Einstein Hilbert action to the zero order and,
we impose the
closure of the fifth coordinate in the field
equations independent of $x_5$, the two
approaches formally overlap.
In fact, the variation of the zero order Lagrangian,
after the integration on the extra coordinate
has been performed coincides without
further restrictions with the Einstein equations.

Thus, the  KK spacetime is a 5D manifold $\mathcal{M}^{\ti{(5)}}$,  direct product $\mathcal{M}^{\ti{(4)}}\otimes \mathcal{S}^{\ti{(1)}}$, between the ordinary 4D spacetime $\mathcal{M}^{\ti{(4)}}$  and  the space like loop  $\mathcal{S}^{\ti{(1)}}$ \cite{Bailin:1987jd,Lb-gr-a}.
The  size on the fifth dimension is assumed to be
below the present observational bound or
\(
L_{\ti{5}}\equiv\int dx_{\ti{5}} \sqrt{-g_{55}}<10^{-18}\rm{cm},
\)
we here refer to this assumption as the ``Compactification hypothesis'',
and the  metrics components  do not depend
on the fifth coordinate (Cylindricity hypothesis): this hypothesis can be  realized assuming we are working in an
effective theory at the lowest order of the
Fourier expansion along the fifth dimension \cite{Overduin:1998pn}.

Moreover it is  assumed that the $(55)$ metric component is a scalar $(g_{55}=-\phi^2)$.
In this way the  evolution of size of the fifth dimension, is entirely controlled by the dynamics of a geometrized real scalar field.
According to the KK dimensional reduction
the 5D line element reads\footnote{With latin capital letters $A$ we
label the 5D indices, where they run in
$\{0,1,2,3,5\}$, Greek and latin indices $a$  run from 0 to 3, the
spatial indexes $(i,j)$ in $\{1,2,3\}$. We consider metric of
$\{+,-,-,-,-\}$ signature. We adopt coordinates $x^{\mu}$ for  the ordinary 4D spacetime while
$x_{\ti{5}}$ is the angle parameter for the fifth circular dimension.}
 as follows:
\begin{equation}\label{cic}
ds^2_{\ti{(5)}}=g_{\mu\nu}dx^{\mu}dx^{\nu}-
\phi^2\left(dx_{\ti{5}}+ekA_{\mu}dx^{\mu}\right)^2,
\end{equation}
$A_{\mu}$ represents the 4D vector potential
and $g_{\mu\nu}$ is the  4D metric tensor; $ek$ is a
dimensional constant such that $e^2k^2=(4G)/c^2$, where $G$ stays for the 4D gravitational constant. In general the KK setup is characterized by the  breaking of the 5D covariance and the 5D equivalence principle \cite{Lb-gr-a}, even if   the translations along the
fifth dimension are allowed to realized the  abelian gauge
invariance of the electromagnetism implemented in the   KK model as a
coordinate transformation in $S^{(1)}$.
\subsection{Particle motion in Kaluza Klein theory: the standard approach}
In the standard, namely ``geodesic'' approach, the matter dynamics in  the KK theories  is generally faced
assuming
 the existence of a 5D point like
test particle  in   $\mathcal{M}^{\ti{(5)}}$.
In \cite{Lacquaniti:2009cr,Lacquaniti:2009rq,Lacquaniti:2009yy} this assumption has been revisited and
the validity of a model with a point like particle in a compactified
dimension is criticized.

The geodesic approach relies on the assumption that the 5D particle  motion is regulated by  a 5D Klein Gordon dispersion relations
$P_{\ti{A}}P^{\ti{A}}=m_{\ti{(5)}}$, where $P_{\ti{A}}$ is the
5 momentum and $m_{\ti{(5)}}$ is  a constant mass parameter associated
to
the particle, in particular the component $P_{\ti{5}}$ is conserved and it defines the particle charge by the
$
q=\sqrt{4G}P_{\ti{5}}\quad\mbox{where}\quad
P_{\mu}P^{\mu}=m_{\ti{(5)}}^2+P_{\ti{5}}^2/\phi^2,
$
the  5D velocities
$\omega^{\ti{A}}$ and the 4D velocities $u^{\ti{A}}$ are defined
respectively as
$
\omega^{\ti{A}}\equiv {dx^{\ti{A}}}/{ds_{\ti{(5)}}}$, $
u^{\ti{A}}\equiv {dx^{\ti{A}}}/{ds}
$,
with $\omega^{\ti{A}}=\alpha
u^{\ti{A}}$ and
$
g_{\ti{A}\ti{B}}\omega^{\ti{A}}\omega^{\ti{B}}=1$, $
g_{ab}u^{a}u^{b}=1
$,
where the $\alpha$ parameter reads $
\alpha\equiv ds/ds_{\ti{(5)}}$, and $ds$ ($ds_{\ti{(5)}})$ states for the 4D (5D) line element.
The dimensional reduction of the equation of motion (the 5D geodesic: $\omega^{\ti{A}}\; {^{\ti{(5)}}}\nabla_{\ti{A}}\omega^{\ti{B}}=0 $)
is recovered, furthermore the cylindric condition  provides a constant of motion in agreement with
the existence of the  Killing vector $(0,0,0,0,1)$.
The  mass parameter $m_{\ti{(5)}}$ is assumed to be
constant. The dimensional reduction
leads to the set
\begin{eqnarray}
\label{Eq:Re_Vi} u^{a}\  ^{\ti{(4)}}\!\nabla_{a}u^{b}&=& e
k\left(\frac{\omega_{5}}
{\sqrt{1+\frac{\omega_{5}^{2}}{\phi^{2}}}}\right)
F^{bc}u_{c}+\frac{1}{\phi^{3}}\left(u^{b}u^{c}-g^{bc}\right)\partial_{b}\phi\left(\frac{\omega_{5}}
{\sqrt{1+\frac{\omega_{5}^{2}}{\phi^{2}}}}\right)^{2},
 \\
\label{Eq:pen_s} \frac{d\omega_{5}}{d s} &=& 0,
\end{eqnarray}
where $F_{ab}=\partial_{a}A_{b}-\partial_{b}A_{a}$ is the Faraday
tensor. Hence, in the standard  scenario  a free 5D test particle   becomes a 4D interacting
particle, whose motion is described by Eqs.\il(\ref{Eq:Re_Vi}).

We note that in particular for
$\omega_{5}=0$ (neutral test particle  case) Eq.\il(\ref{Eq:Re_Vi})
becomes a geodesic one. Moreover, even  in a free electromagnetic
scenario, the charged
particles  (with $\omega_{5}\neq0$) do not follow in general a geodesic motion, being coupled with the extra dimensional scalar field by
a $\omega_{5}$ function.
There is in fact in Eq.\il(\ref{Eq:vol_a}) a force term determined by the electromagnetic field  (no more geometrized
 in the 4D scenario reduced from the 5D KK model) and by the  scalar field dynamics. Any variation of the fifth dimension size appears as  a force on the particle motion in the 4D Universe. With a constant scale factor the last term on the right side in Eq.\il(\ref{Eq:Re_Vi}) vanishes.
The coupling factors
are
 functions of $\omega_{5}$, in particular the
electrodynamic (EM) coupling factors, in terms of the effective particle
 charge to mass ratio $q/m_{\ti{(5)}}$ is
\begin{equation}\label{Eq:vol_a}
\frac{q}{m_{\ti{(5)}}}=ek \frac{\omega_{5}}
{\sqrt{1+\frac{\omega_{5}^{2}}{\phi^{2}}}}.
\end{equation}
Noticeably the right side of  Eq.\il(\ref{Eq:vol_a}) is in general no constant
and  it is always upper bounded. Even  if one sets $\phi=1$, that is within the assumption that the extra dimension has a constant size,  it is
$q<m_{\ti{(5)}}$ which is unacceptable for every known elementary
particle. This problem is related to  the huge
massive mode of the KK tower
(\cite{Lacquaniti:2009cr,Lacquaniti:2009rq,Lacquaniti:2009yy}).
The alternative dynamic model, discussed   in the next section seems to solve in particular this problem.
%
\section{Alternative scenario: the  Papapetrou's approach}\label{Sec:papaint}
\subsection{Papapetrou approach  to the particle dynamics in  the Kaluza Klein model}\label{Sec:papa-mo_pa}
In \cite{Lacquaniti:2009cr,Lacquaniti:2009rq,Lacquaniti:2009yy} a new proposal for the  matter dynamics in the KK model is given,  adopting a
Papapetrou multipole expansion \cite{papapetrou} within the compactification hypothesis: extending  the cylindricity hypothesis to a generic  5D matter tensor  it is assumed that  the particle is
described as a localized source in $M^{\ti{(4)}}$ but still
delocalized along the fifth dimension. This hypothesis
 relays on the idea that, since the fifth dimension
is compactified to Planckian like scales,
it makes no sense to handle with a classical
fifth component of the particle velocity and therefore
this leads to the treatment of the matter sources
through the energy momentum description. Such a
point of view leads to adopt a Papapetrou scheme
\cite{papapetrou} to fix the dimensionally reduced
equation of motion for fields and the  matter.
Performing   the dimensional reduction either for the metric fields and
 the matter   it is provided  a consistent approach  that removes the  problem of huge massive
modes, hence a consistent set of equations is wrote\cite{Lacquaniti:2009wc}.

First a 5D energy momentum tensor $^{\ti{(5)}}\!\mathcal{T}^{\ti{AB}}$ is
associated to the generic 5D matter distribution governed by the
conservation law $^{\ti{(5)}}\!\nabla_{\ti{A}} \
^{\ti{(5)}}\!\mathcal{T}^{\ti{AB}}=0$ and not depending on the fifth
coordinate, thus $\partial_{5} \
^{\ti{(5)}}\!\mathcal{T}^{\ti{AB}}=0$ (here $^{\ti{(5)}}\!\nabla$
 is the covariant derivative compatible with
the 5D metric).
Performing a multipole expansion of  $
^{\ti{(5)}}\!\mathcal{T}^{\ti{AB}}$,  centered on a trajectory
$X^{\alpha}$,  the lowest order of the procedure gives the equation
of motion for a test particle:
\bea\label{Eq:pa_ma}
u^{\mu}\,^{\ti{(4)}}\nabla_{\mu}u^{\nu}=\frac{q}{m}
F^{\nu\rho}u_{\rho}+A(u^{\rho}u^{\nu}-g^{\nu\rho})\frac{\partial_{\rho}\phi}{\phi^3},
\quad
 \frac{\partial m}{\partial
x^{\mu}}=-\frac{A}{\phi^{3}}\frac{\partial\phi}{\partial x^{\mu}}.
\eea
The coupling factors  are the charge $q$, coming from the current vector   and the scalar charge $A$
as follows,
\begin{eqnarray}
m &=& \frac{1}{u^{0}}\int d^{3}x \sqrt{-g_4} \phi  \
^{\ti{(5)}}\!\mathcal{T}^{00},\quad \phi \sqrt{-g_4} T^{\mu\nu}=\int ds
m \delta^{4}\left(x-X\right) u^{\mu} u^{\nu},
\\\label{qdef}
q &=& e k \int d^{3}x \sqrt{-g_4} \phi  \
^{\ti{(5)}}\!\mathcal{T}_{5}^{0},\quad e k\phi \sqrt{g}
T^{\mu}_{5}=\int ds q\delta^{4}\left(x-X\right) u^{\mu}= \sqrt{-g_4}
J^{\mu},
\\\label{Adef}
A &=& u^{0} \int d^{3}x \sqrt{-g_4} \phi
^{\ti{(5)}}\!\mathcal{T}_{55},\quad\phi \sqrt{g} T_{55}=\int ds A
\delta^{4}\left(x-X\right),
\end{eqnarray}
these relations provide also the definitions for
coupling factor $m$, $q$, $A$ and the according definitions for the effective test particle tensor
  $T^{AB}=\mathcal{T}^{AB}l_{\ti{(5)}}$ and $e{k}$ reads $(e{k})^2={4G}/{c^2}$, where
$g_4$ is the determinant of the 4D metric\cite{Pugliese:2011um,Lacquaniti:2010zz}.
The Eq.\il(\ref{Eq:pa_ma}) describes the   motion of a
point like particle of mass $m$ in the 4D spacetime, coupled to the
electromagnetic field through the charge $q$ and to the scalar field by
the new quantity $A$.
The continuity equation $\apis{4}\nabla_{\mu}\apis{4}J^{\mu}=0$, derived
within the procedure itself, implies that charge $q$ is conserved.
The parameter $m$ correctly represents the mass of the particle,
which turns out to be localized just in the ordinary 4D space, as it
is envisaged by the presence of a 4D Dirac delta function in the
above definitions. It can be proved that the KK tower of massive modes is
suppressed, and the $q/m$ ratio is no more upper bounded. Indeed, it
can be proved that the motion of the particle is correctly governed
by a dispersion relation of the  form
$
P_{\mu}P^{\mu}=m^2\,,
$
where  here and in the following the dimensional
index $\apis{4}$ is dropped.
 Nevertheless, in general as a consequence of the
coupling to the scalar field, the   mass is now variable
and it is precisely related to the variation of the
scalar field and the new coupling $A$ (which has a pure
extra dimensional origin) along the path. An interesting scenario
concerning $A$ to be investigated is  $ A\,\infty
\,m\phi^2$: by this way Eq.\il(\ref{Eq:pa_ma}) admits an easy
integration, providing a power law dependence of the mass on the scalar
field and, more important, restoring the particle free falling universality
  when a vanishing
electromagnetic field is considered,
while the requirement $A = 0$, implies that the  mass $m$ is constant.
 \subsection{Kaluza Klein field equations}\label{sec:Papapetroumatter}
The revision  of the consequences of the  dimensional compactification  in the KK spacetimes
has finally  led to the  reformulation of the field  equations in the matter in this revisited KK context.
The full system of the 5D Einstein equation in presence of the  5D matter described by a   5D matter tensor $\mathcal{T}^{AB}$ is:
\be
^5G^{AB}=8\pi G_5 \mathcal{T}^{AB},
\label{kkk}
\ee
where the  5D Bianchi identity holds,
while from the cylindricity hypothesis concerning the matter field it is
$
\partial_5 T^{AB}=0,
$
where $G_5$ is the 5D Newton constant,
 $T^{AB}=\mathcal{T}^{AB}l_{\ti{(5)}}$, and  $G=G_5l_{\ti{(5)}}^{-1}$, where the coordinate length of the extra dimension $l_{\ti{(5)}}=\int dx^5$.
The components $T^{\mu\nu}$, $T_5^{\mu}$, $T_{55}$ are a 4D tensor, a 4D vector and a scalar respectively.
Introducing the quantities
$
T^{\mu\nu}_{\ti{matter}}=l_{\ti{(5)}} \phi  \mathcal{T}^{\mu\nu}=\phi T^{\mu\nu}$, $j^{\mu}=ek\phi T_5^{\mu}
$,
and $\vartheta=\phi T_{55}$, we have:
\be
\nabla_{\mu}j^{\mu}=0,
\ee
that introduces a conserved current $j_{\mu}$, related to the U(1) gauge symmetry and  coupled to the tensor $F_{\mu\nu}$, together with  the  conservation equation for  $T^{\mu \nu}_{\ti{matter}}$
\be
\nabla_{\rho}\left(T^{\mu\rho}_{\ti{matter}}\right)=-g^{\mu\nu}\left(\frac{\partial_{\rho}\phi}{\phi^{3}}\right)\vartheta+F^{\mu}_{\phantom\
\rho}j^{\rho},
\ee
coupled to the field $\phi$ and $A_{\mu}$ by the matter terms $T_{55}$ and $j_{\mu}$ and representing the energy momentum density of the ordinary 4D matter.

In the limit  $\phi=1$ we recover the conservation law for an electrodynamic system, and setting   $T_{55}=j^{\mu}=0$ it is $\nabla T^{\mu\nu}_{\ti{matter}}=0$.
Taking now into account these definitions, the reduction of Eq.\il(\ref{kkk}) leads to the set
\bea\label{ki}
G^{\mu\nu}&=&\frac{1}{\phi}
\left(\nabla^{\mu}\partial^{\nu}\phi+8\pi T^{\mu\nu}_{em} G -g^{\mu\nu}\Box\phi +8\pi G T^{\mu\nu}_{\ti{matter}}\right),\\
\label{kl} \Box\phi&=&\frac{8 \pi}{3}G
\left(T_{\ti{matter}}+2\frac{\vartheta}{\phi^2}\right)-G\phi^{3} F^{\mu\nu}F_{\mu\nu},
\eea
for the Einstein and KK field,
where
\be\label{k12}
\frac{R}{2}=-\frac{3\phi^3(ek)^2F^{\mu\nu}F_{\mu\nu}}{8}
+8\pi G \frac{\vartheta}{\phi^3},
\ee
the KK Maxwell equation is now
\be
\nabla_{\nu}\left(\phi^3F^{\nu\mu}\right)=4\pi j^{\mu}.
\ee
In such a scenario the function $\vartheta$   is still undetermined and it should  be fixed for a given background, while the reduced field equations correctly  describe an Einstein Maxwell system and  the full KK dynamics in presence of  the matter source terms.  The model function $ \vartheta $ plays a similar role to the model  parameter $A$ in the equations for the test particle dynamics.
With respect to the pure Einstein Maxwell system there are now  two additional scalar fields: the usual KK scalar field, $\phi$, plus a scalar source term, $\vartheta\equiv l_{\ti{(5)}} \phi T_{55}$.
\section{Generalized Schwarzschild spacetime}\label{Sec:GSS}
In this section we introduce a special family of metrics, solutions of the  electromagnetic free 5D vacuum KK equations
(\cite{Gross:1983hb,sorkin,davisonowen}).
This metric family has many remarkable properties: firstly its 4D counterpart  is spherical symmetric (4D spherical symmetry) and it is time independent: the ordinary
4D spacetime $M^{\ti{(4)}}$ of the direct product $M^{\ti{(4)}}\otimes
\mathcal{S}^{\ti{(1)}}$ is spherically symmetric; in other words the
sections $t = {\rm{const}}$, $r = {\rm{const}}$ and $x_{\ti{5}} = {\rm{const}}$ of
$M^{\ti{(5)}}$ are $S^{\ti{(2)}}$ (spherical surfaces in the ordinary
3D space).
In addition, for suitable limits on the family parameters these solutions can be reduced in their 4D counterpart to the Schwarzschild  metric.
For this reason  this set of solutions is also known in the literature as Generalized Schwarzschild spacetime (GSS).
This is a particularly interesting and versatile case to study because of  its symmetric properties, and the amenability to a  known 4D spacetime scenario   widely studied in the astrophysics of the 4D Universe,
\begin{figure}[h!]
\centering
 \includegraphics[width=0.57\hsize,clip]{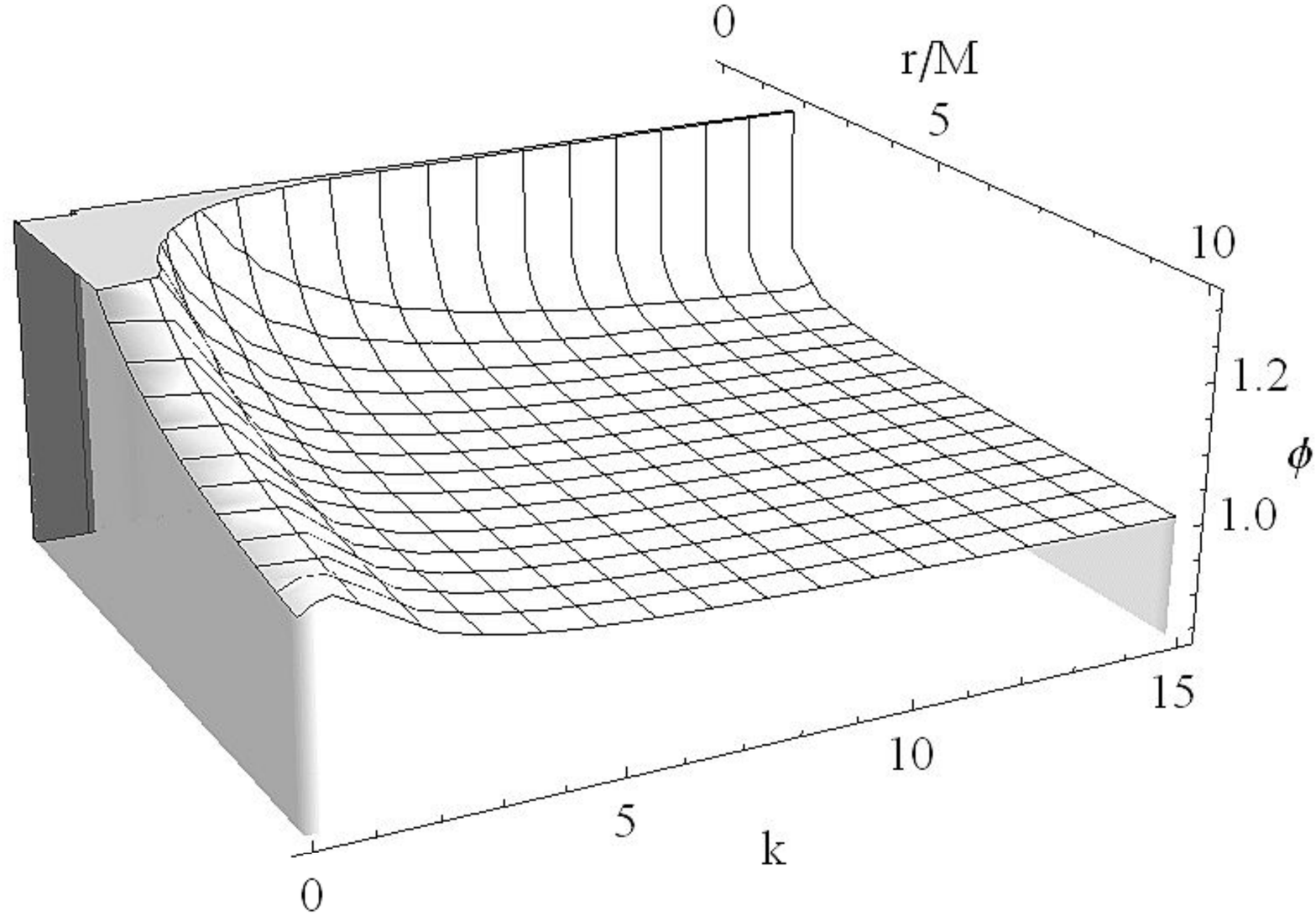}
\caption[font={footnotesize,it}]{\footnotesize{The extra dimensional scale factor $\phi=\sqrt{-g_{55}}$ as function of $r/M$ and $k$.}
}\label{Fig:d_u_l}
\end{figure}
The one parameter family  of metrics is
\begin{equation}\label{Eq:C_G_M}
ds_{\ti{(5)}}^{2}=\Delta(r)^{\epsilon k}dt^{2}-\Delta(r)^{-\epsilon
(k-1)}dr^{2}-r^{2}\Delta(r)^{1-\epsilon (k-1)}d\Omega^{2}-
\Delta(r)^{-\epsilon }dx_5^{2}\, ,
\end{equation}
in the 4D spherical polar coordinate\footnote{Consider
$t\in\Re$, $r\in\left]2M,+\infty\right]\subset\Re^{+}$,
$\vartheta\in\left[0,\pi\right]$, $\varphi\in\left[0,2 \pi\right]$}
$\left\{t,r,\theta,\varphi\right\}$ where
$d\Omega^2\equiv\sin^2\theta d\varphi^{2}+d\theta^{2}$,
where $\Delta(r)=\left(1-2M/r\right)$, with $M$ is a constant.
Each values of the parameter $k\geq0$ sets a specific metric of the family,  the parameter $\epsilon\geq0$ is in fact constrained by
$ \epsilon^{2}\left(k^{2}-k+1\right)=1 $. For a
review about the  constraints on the $k$ parameter see  \cite{Overduin:1998pn} and  \cite{Lacquaniti:2009wc}.
The GSSs  are asymptotically flat and,  for each finite value of $k$, GSS has a naked singularity situated in $r=2M$, that resolves in a black hole
 only in the Schwarzschild limit for $(\epsilon,k)$.
The
Schwarzschild limit is recovered on the spacetime section $dx_{\ti{5}}=0$ for $\epsilon\rightarrow0,\;
k\rightarrow\infty$.

Moreover,  as $r$ approaches $2M$, $g_{tt}$ reduces to zero  while $g_{55}$ explodes to infinity, therefore
the  length of the extra dimension increases as well as $r$
approaches $2M$, Fig.\il(\ref{Fig:d_u_l}).

These solutions represent extended  objects:
the GSS naked singularity is surrounded by the induced scalar matter with a trace free energy momentum tensor.

The constant parameter $M$
is  related to the mass of a central body which is supposed to act as source of the gravitational field.
Noticeably, the Schwarzschild limit is obtained when $\phi=1$, and in this limit $M$ is the Schwarzschild mass.
In fact,{  the total mass of this configuration has been calculated with many different  definitions  (see for example \cite{PoncedeLeon:2007bm}), we refer here to the gravitational mass $M_g$,  inside a 3D volume,  as given by the Tolman-Whittaker
formula:
\(
M_g=\int(T^0_0-T^1_1-T^2_2-T^3_3)\sqrt{-g_4}dV_3
\)
that can be written in isotropic coordinate as $M_g(r)=(2\epsilon k/a) [(ar-1)/(ar+1)]^{\epsilon}$ where $a$ is a constant related to the mass of the central body, (see \cite{Overduin:1998pn} and references  therein). }
This is, in the Schwarzschild limit, $M_g=M$; while, for each finite value of $k$, it  is $M_g=\epsilon
kM$ at infinity and  it goes to zero as $r$ closes the singularity and $dV_3$ is the ordinary
spatial 3D volume element.

Hence, for every finite values of $k$, GSSs are  naked
singularities surrounded by the induced scalar matter, and only in the limit
$k\rightarrow\infty$ they are black holes with an horizon in $r=2M$
and vacuum for $r>2M$.

But, despite
of the naked singularity feature showed far from the Schwarzschild
limit, GSSs  are supposed to describe in principle the
exterior spacetimes of any astrophysical sources characterized by the
required  symmetries,  embedded
in a cloud of a real scalar field.
 This consideration gave rise to the work in \cite{Pugliese:2011yh} where an internal solution of the 5D KK equations was found as model for   a stellar object in the 4D Astrophysics.
Here we consider GSS with  $k>1$ as  the (5D vacuum) spacetime, surrounding a compact object with a radius
$R>2M$.

In the attempt to provide any possible constraints for the $k$ parameter many efforts   have been made to simulate known astrophysical objects  with a metric (\ref{Eq:C_G_M}),
ie it was attempted to identify a spacetime generated by the selected  source  (for example the solar system) with a certain family  metric,  associating the gravitational source  to a specific $k$, \cite{Overduin:1998pn,Xu:2007dc,PoncedeLeon:2006xs,PoncedeLeon:2007bm}.
However
modelling the Sun by a GSS,    requires a fine tuning of
the characteristic parameter
\cite{Overduin:2000gr} and in any case  the various works and estimates associated with this model show that extra dimensions should play  a negligible role in the solar
system dynamics.
The tests elaborate in \cite{Lacquaniti:2010zz,Lacquaniti:2009rh,Pugliese:2011um,Pugliese:2011yh,Lacquaniti:2009wc}, and briefly discussed  in the following sections, within the Papapetrou approach stay as a valid  alternative to these studies.
\section{Test particle dynamics in the Generalized Schwarzschild
spacetimes: the Papapetrou analysis}\label{Sec:CircularorGSS}
In this section we illustrate the main results concerning the time like circular orbits in the GSS
within the Papapetrou approach.
Considering  the 4D momentum
 $P_{\mu}=m u_{\mu}$ the mass $m$ is
a function of the radial coordinate $r$,
constant along the circular orbits (at fixed $r$). In the
Schwarzschild limit, or asymptotically, where $\phi=1$ we have
$m=m_0=const$. It is always possible to build the  constants
of motion $E$ and $L$ defined as follows:
\begin{equation}\label{WIWA}
E\equiv\xi^{a}_{(t)}P_{a}=m g_{tt} u^{t}, \quad L\equiv\xi^{a}_{(\varphi)}P_{a}= m
g_{\varphi\varphi} u^{\varphi},
\end{equation}
where $\xi^{a}_{(t)}$ and $\xi^{a}_{(\varphi)}$ are  metric Killing
fields,
thus  $E$ and $L$, functions of $r$ and $k$, are interpreted as the energy at infinity of
the particle and the  total angular momentum, respectively.

Adopting the standard procedure we define an \emph{effective potential}
\begin{equation}\label{Eq:ma_spe_al}
V_{eff}\equiv E=
\sqrt{g_{tt}\left(m^{2}-\frac{L^{2}}{g_{\varphi\varphi}}\right),
}\end{equation}
describing the motion of a  test particle  of mass $m$  in a circular orbit around the source in terms of the  1D motion of a particle in an effective potential $ V_{eff} $,  \cite{MisThoWhe73}. This function of the position $ r $ and the orbital angular momentum $ L $, takes into account the centrifugal effects, through the angular momentum, and the gravitational field through the explicit dependency on the metric components $ g_{tt} $ and $ g_{\varphi \varphi} $. However, the dependence on the fifth dimension through the scalar field is enclosed by the mass factor   $ m $ that appears explicitly in Eq.\il(\ref{Eq:ma_spe_al}), this term, as  seen in Section\il(\ref{Sec:papa-mo_pa}), also depends on the particular KK model beyond the choice of  the parameter $ A $.

The orbital angular momenta and the particle  energy in terms of the orbital radius $ r $, can be obtained by solving the equation
\be\label{E:circ_orbi_:def_1}
V_{eff}'=0, \quad V_{eff}=E.
\ee
Equation (\ref{E:circ_orbi_:def_1}) can also be solved by the radius $r$ in terms of the orbital momentum $L$. These radii will  generally depend on the intrinsic properties of the  source, and on the particular adopted Papapetrou model   ($A$), thus  in this case   it will be $r = r (M, k; A)$.
As in the  Schwarzschild case  a set of  particular orbits  exist  that determine the stability regions
for circular orbits.  Specifically, this is the\emph{last circular orbit radius} $r_{lco}$ and the \emph{last stable circular orbit radius} $r_{lsco}$. These correspond  to the maximum and minimum of the effective potential as a function of $r$ respectively. They  are solution of Eq.\il(\ref{E:circ_orbi_:def_1}) with $V_{eff}''<0$ for  $r_{lco}$ and
$V_{eff}''>0$ for  $r_{lsco}$.

In general  one can say that in the region   $r<r_{lco}$ there are no circular orbits, while circular  orbits in  $r_{lco}<r<r_{lsco}$  are unstable, and finally all orbits with {$r>r_{lsco}$} are stable.
Clearly  one can expect that these particular orbits and the regions of orbital existence and stability are deformed respect to the  orbital regions in the 4D Schwarzschild case. Each GSS  is  in general characterized by a couple of  different $(r_{lco}, r_{lsco})$.
However it is possible to show that
$r_{lco}/M=\left[1+\epsilon(2k-1)\right]$ turns to be a function of  $k$ only.
In \cite{Lacquaniti:2010zz}  different  scenarios  for the  test particle circular orbits
were addressed  considering  the cases $A=0$,  $A=$constant and
$A=\beta m \phi^2$,  where $\beta$ is a real number. In the next sections we will outline the main results  of this research  showing in particular  the deformation of the orbital stability regions respect to  the 4D Universe model.
\subsection{The case  $A=0$}
In the KK scenario  defined  by   $A=0$, all the  charged and  neutral test particles in the GSS follow a geodesic equation
\begin{equation}\label{Eq:PapA0}
u^{\mu}\,^{\ti{(4)}}\nabla_{\mu}u^{\nu}=0,\quad \mbox{with}\quad \partial_\mu m=0,
\end{equation}
 in the ordinary
4D spacetime,  where $m$ is a constant and  no
scalar field coupling term appears.  {The quantities with subscript $ (_0) $, identify here and hereinafter the constant quantities  determined by the initial conditions.}

Studying the potential $V_{eff}$ as function of the orbit radius $r$,
we find the particle energy ${E}$ and
the angular momentum $L$    of  the circular orbit
\cite{Lacquaniti:2009wc}.
We recover the
last circular orbit radius   and the last stable circular orbit radius
\begin{eqnarray}\label{Drr}
r_{lco}=\left[1+\epsilon(2k-1)\right]M,
\quad r_{lsco}=\left[1+\epsilon(3k-2)+\epsilon\sqrt{(k-1)(4 k-1)}\right]M\, .
\end{eqnarray}
function of $k$.
It is    $r_{lco}<3M$ and  it reduces to the Schwarzschild limit  $r_{lco}=3M$ for large $k$
\cite{Lacquaniti:2009wc,Lacquaniti:2009yy}. However,  at the decreasing of
$k$, the last circular orbit radius approaches the
point $r=2M$  in particular  for $k=1$ it is $r_{lco}\equiv2M$.
On the other hand
 it is
$r_{lsco}<6M$ $\forall k>1$ and for $k=1$ we have $r_{lsco}=2M$ while in the Schwarzschild's limit we have $r_{lsco}=6M$.

Thus, in this case the analysis of  the  circular orbits and their stability has  provided a significant constraint on the possible GSS   and the physics of the 4D reduced Universe   by the  KK  Papapetrou procedure.
Indeed,  the orbital region where the circular motion  is allowed and   the orbital region in which these  orbits are stable are in general larger then the  analogous regions in Schwarzschild spacetime   as much as the reduced solution deviates from its Schwarzschild limit. Fig.\il(\ref{Fig:star_t})-left shows the radii  $ r_{lsco} $ and $ r_{lco} $ as functions of $ k $ and their asymptotic limits, the energy
${E}_{lsco}$  and the particle angular orbital momentum  ${L}_{lsco}$ in the last stable circular orbit are plotted.
It is   $E_{lsco}<{2
\sqrt{2}}/{3}m=0.942809m$ its Schwarzschild's limit for all values of $k\gtrsim 1$, while ($L_{lsco}=2\sqrt{3}Mm=3.4641Mm$) Schwarzschild's limit  value for $k>3.45644$,
 Fig.\il(\ref{Fig:star_t}). However, as noted in \cite{Lacquaniti:2009wc} this fact should not be read as a direct
consequence of a possible motion along the fifth dimension, since
Eq.\il(\ref{Eq:PapA0}) does not depend on it, neither on
$g_{\ti{55}}$. We interpret this fact  as a feature
related to the deformation of the Schwarzschild metric as long as $k$ is
sufficiently small and this  seems to be
confirmed also by the fact that Eqs.\il(\ref{Eq:PapA0}) are
the same that  obtained from the geodesic approach with
$\omega_{5}=0$.
\subsection{The case {$A=const$}}
In the  model $A={\rm{constant}}\neq0$  the  particles follow the curves
\begin{equation}\label{Eq:Ba_a}
u^{a}\ ^{\ti{(4)}}\!\nabla_{a}u^{b}=
(u^{b}u^{c}-g^{bc})\left(2\frac{\partial_{c}\phi}{\phi}\right)\,\quad
m=\frac{A}{2\phi^{2}}+m_{0}-\frac{A}{2\phi_{0}^{2}}\,.
\end{equation}
Note that in this case the mass is in general a function of the adopted model  through the selection of the parameter $ A $ and indirectly function of the orbital radius  through the scalar field $ \phi $, that regulates the size of the fifth dimension. However in the Schwarzschild limit it is $m=m_{0}$ and  if we set   $A= 2 m_0\phi_{0}^{2} $ it is  $m= A/2\phi^2$.

\begin{figure}[h!]
\centering
\includegraphics[width=0.35\hsize,clip]{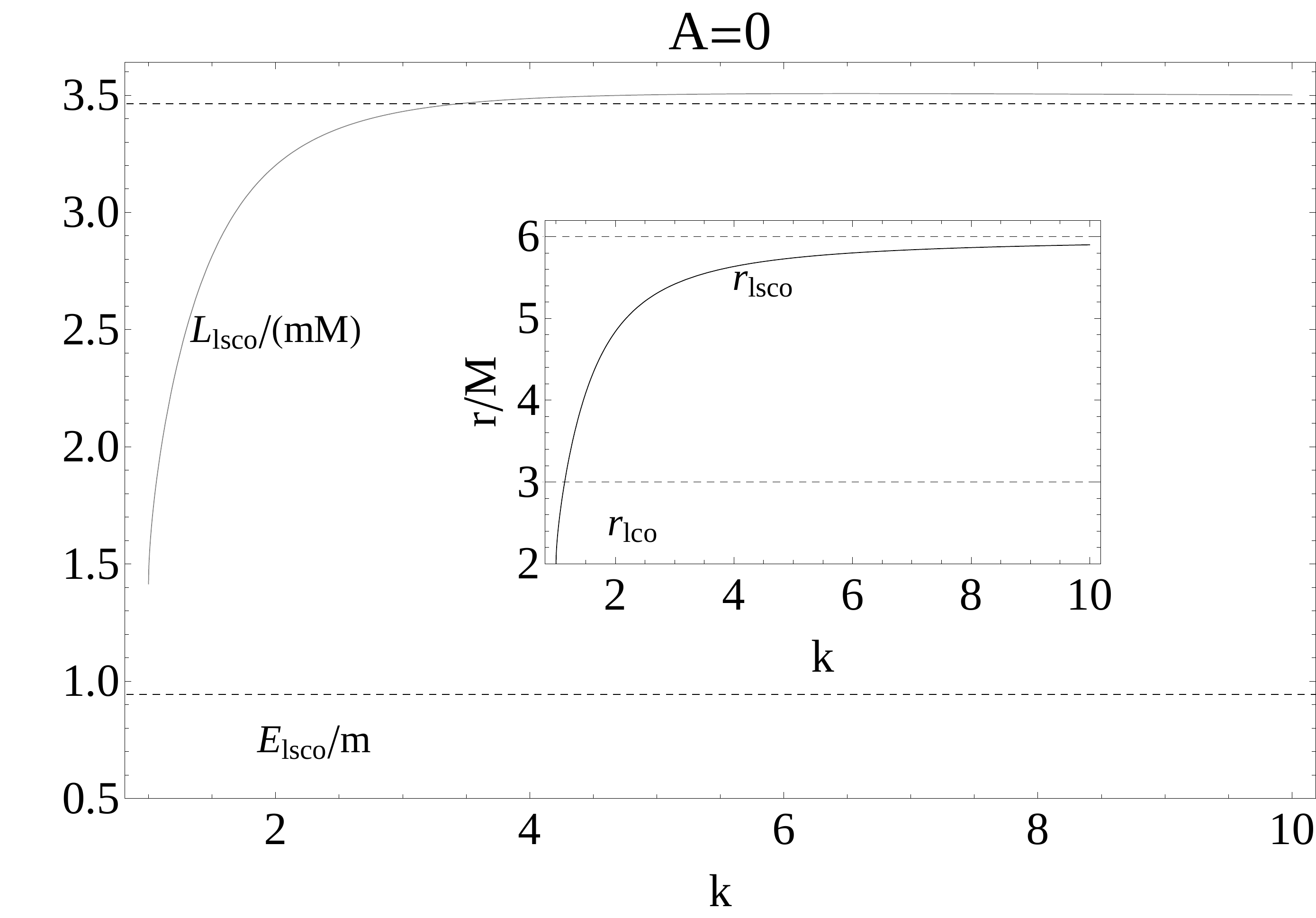}
\includegraphics[width=0.35\hsize,clip]{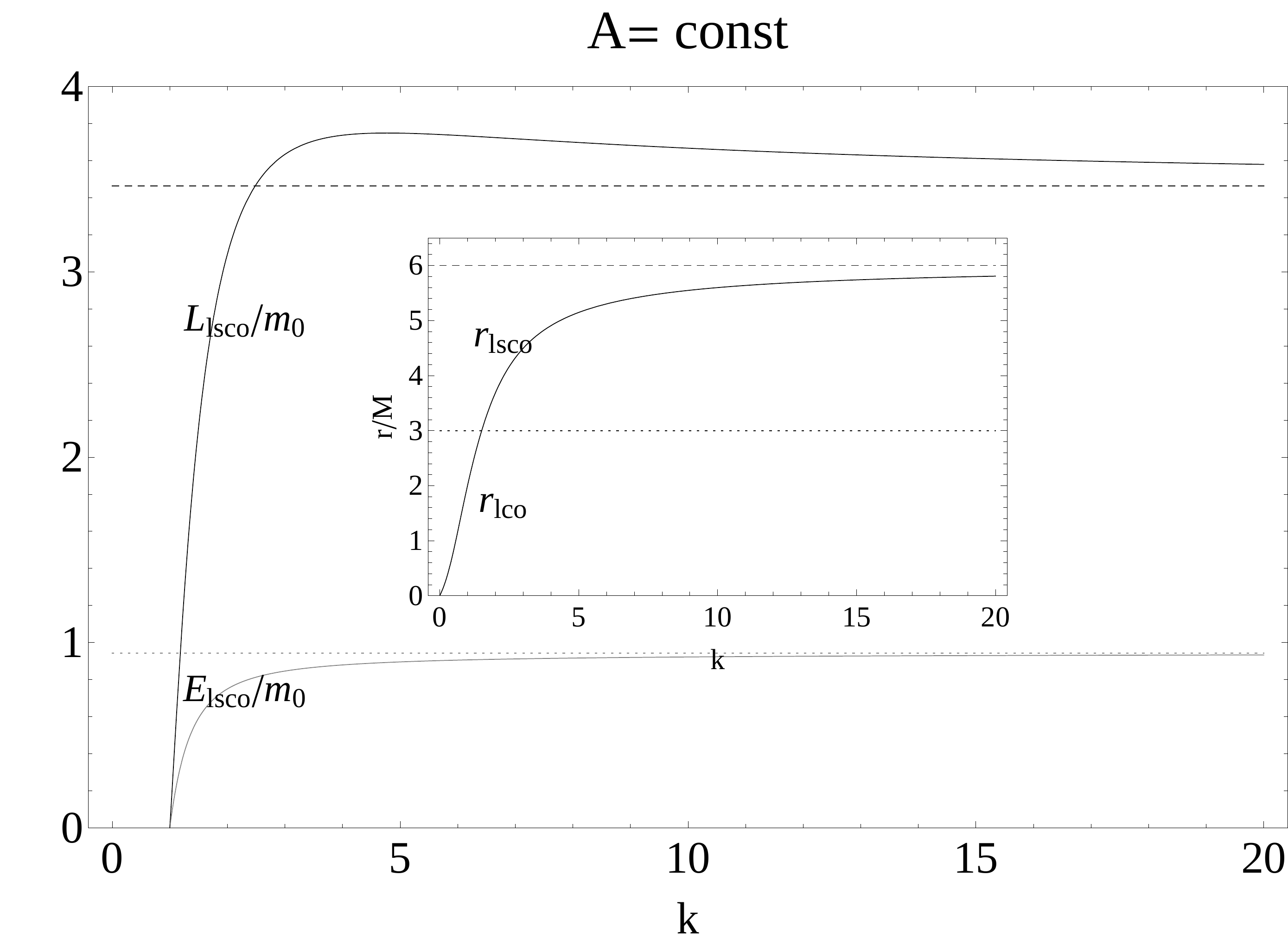}
\includegraphics[width=0.35\hsize,clip]{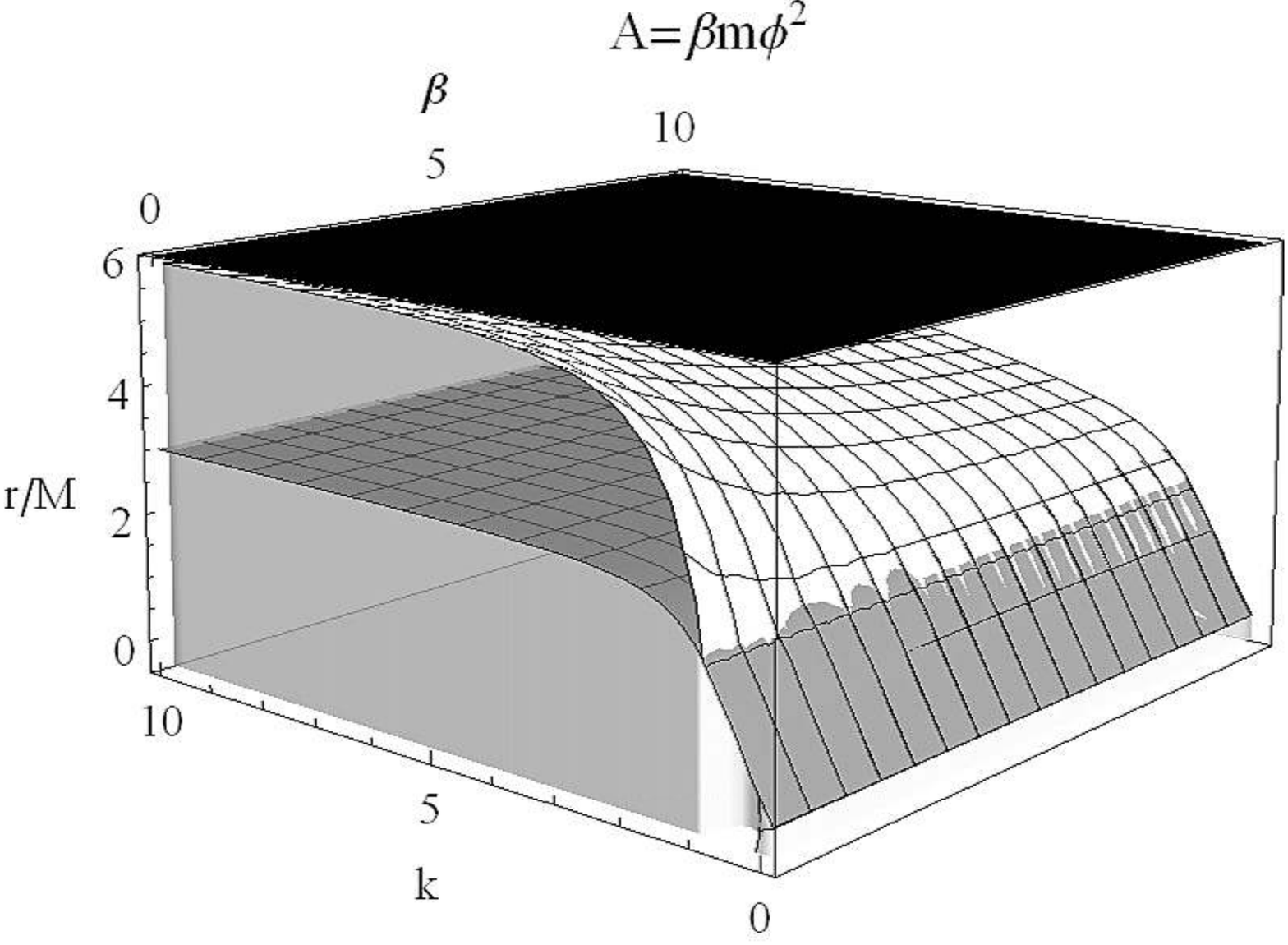}
\caption[font={footnotesize,it}]{\footnotesize{Last circular orbits radius $r_{lco}/M$  (gray
line) and    last \emph{stable} circular orbit radius $r_{lsco}/M$
(black line) as functions of the $k$ parameter are plotted, for $A=0$ (left plot) and $A=$constant (right plot). Inside plot: the ${E}_{lsco}/m$ (black line) and
${L}_{lsco}/m$ (gray line) are
plotted as functions of k. Schwarzschild's limits for the
energy, the angular momentum and the orbits are also plotted (dashed lines). Center plot:
case $A=\beta m \phi^{2}$  last circular orbits radius $r_{lco}/M$  (gray.
surface) and    last \emph{stable} circular orbit radius $r_{lsco}/M$ (light gray surface) as functions of $\beta$ and $k$. The black plane is $r_{lsco}=6M$.}} \label{Fig:star_t}
\end{figure}

\begin{figure}[h!]
\centering
\includegraphics[width=0.5\hsize,clip]{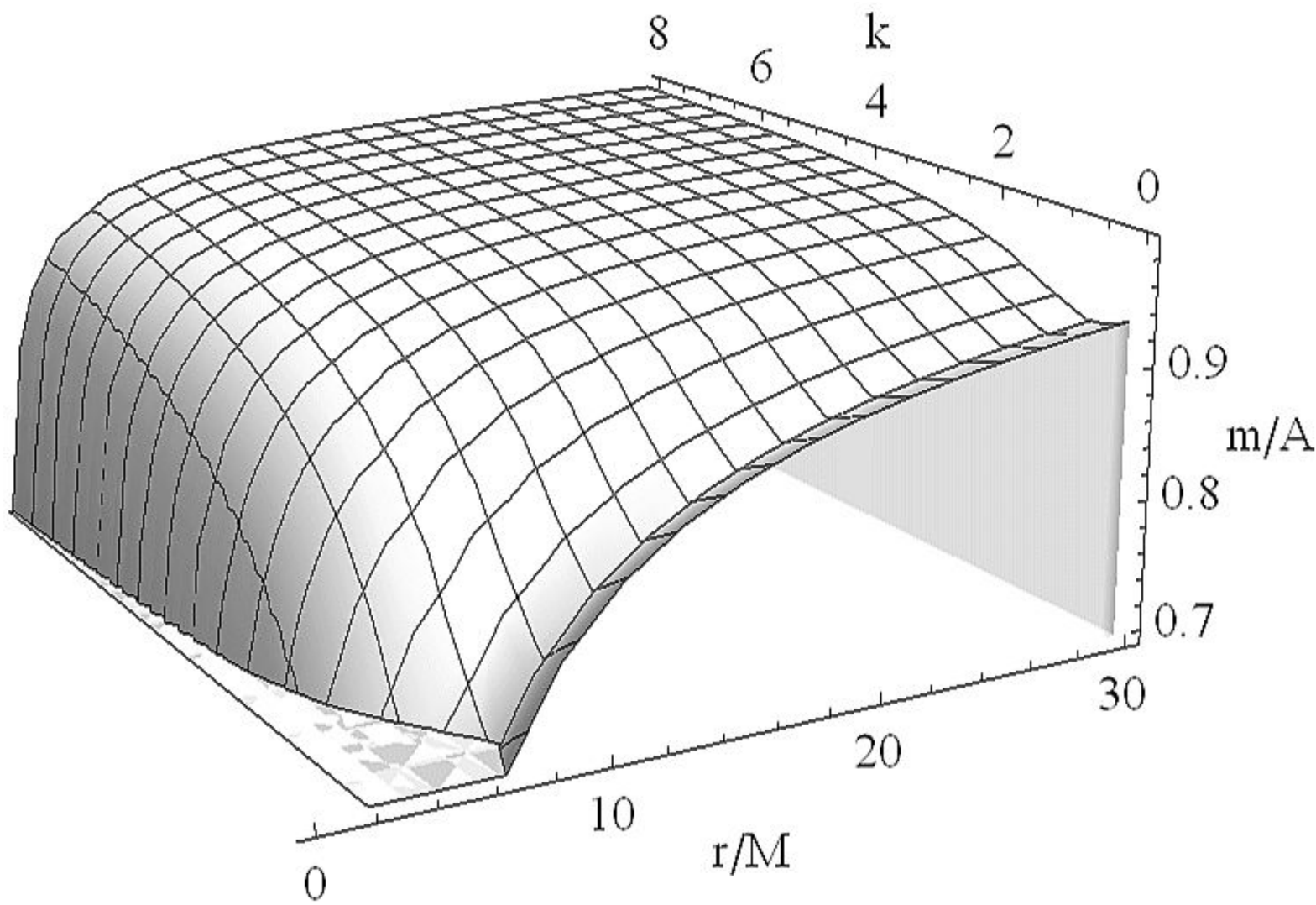}
\caption[font={footnotesize,it}]{\footnotesize{Case: $A=$constant$\neq0$. The mass $m/A$ as function of $k$ and $r/M$.}}\label{Fig:mma}
\end{figure}

From the effective potential we find:
 \begin{equation}\label{Eq:For_0090}
r_{lco}\equiv M
 \left[1+\epsilon\left(2k -1\right)\right]\quad r_{lsco}\equiv\frac{\sqrt{M^{2}\left[4+(15k-8)
\epsilon^{2}+5(8-3k)\epsilon
^{4}\right]}+M\left[3+\epsilon(2+k-11\epsilon+5k\epsilon)\right]}{(2+k)\epsilon}\,.
\end{equation}
It should be noted that  the orbital radii in Eq.\il(\ref{Eq:For_0090}) do not  explicitly depend on $A$, Fig.\il(\ref{Fig:star_t})-center.
As for the model $A=0$ also in this case it is $r_{lsco}<6M$ for each $k$.   {Nevertheless we note that in the case $A\neq0$,
the energy $E_{lsco}/m_{0}$
and the  momentum $L_{lsco}/(m_{0}M)$   are considered as functions of  $m_0$ only. The mass $m/A$ has been plotted in Fig.\il(\ref{Fig:mma})}, as function of $r/M$ and $k$.

\subsection{The case $A=\beta m \phi^{2}$}
Finally we set $A=\beta m \phi^{2}$, with $\beta\in \Re$. In this model the test  particles  will follow the curves
\begin{equation}\label{Eq:ga_after}
u^{a}\ ^{\ti{(4)}}\!\nabla_{a}u^{b}=
\left(u^{b}u^{c}-g^{bc}\right)\frac{\partial_{c}\phi}{ \phi}\beta,
\quad m=\frac{m_{0}\phi_{0}^{\beta}}{\phi^{\beta}},
\end{equation}
as for the case $A=$constant$\neq0$ the mass $m$ is no more a constant,
but here  $m_{0}\phi_{0}^{\beta}=const$.
The $A=$constant case  is recovered with
 $\beta=2$ and  $A^{2}=4 m_{0}^{2} \phi_{0}^{2\beta}$.
With the parameters $A^{2}\equiv m_{0}^{2} \phi_{0}^{2\beta}$,
where $A>0$ and $\beta=2n$ with $n\in\mathbb{Z}$. In \cite{Lacquaniti:2010zz} it was shown that
for $k>-\beta$
$$ r_{lco}\equiv M
\left[1+\epsilon\left(2k +1\right)\right],
$$
$$
r_{lsco}\equiv
M\frac{3+\epsilon[k+\beta+(-3+k+2k\beta-\beta(2+\beta))\epsilon]}{(k+\beta)\epsilon}
+
$$
$$
+M\frac{\sqrt{4+\epsilon^{2}\left[-3k(1+2\beta)\left(\epsilon^{2}-1\right)+(2+\beta)\left(\beta-4
+\left(\beta^{3}+2\right)\epsilon
^{2}\right)\right]}}{(k+\beta)\epsilon},
$$
and $r_{lco}<3M$, notice that in this case the last
stable circular orbit  radius    depends on the  two free parameters, $k$  and $\beta$. Moreover  $r_{lsco}<6M$ for $\beta>0$,
while for $\beta<0$ and $k>-\beta$,  $r_{lsco}>6M$ is possible.
For $\beta=2$ we recover the same physical situations sketched in
the case $A=0$.
Fig.\il(\ref{Fig:star_t})-right plots $r_{lsco}$ and  $r_{lco}$ as function of $(k,\beta)$ the planes of their asymptotical limits are also shown.  We note that in general $r_{lsco}<6M$ and  the difference $|r_{lsco}-6M|$  increases with $\beta$.
\section{Stars in five dimensional Kaluza Klein theory}\label{Sec:KK-Stars}
In \cite{Lacquaniti:2009rq}  a 5D interior solution of the KK equations was found
that matches the GSS at some radius $r=R$.
The existence of such configuration allows to remove the naked singularity that characterizes the GSS. This solution has been interpreted as a 5D stellar model (say a KK star) with a perfect fluid coupled with scalar field such that  a cloud of real scalar field surrounds the 4D object.
This solution has been used as the source of the gravitational field in \cite{Pugliese:2011um}, in the analysis of the EM emission spectrum of a test  particle  radially infalling in this spacetime towards the surface of this stellar object.

\begin{figure}
\includegraphics[width=0.325\hsize,clip]{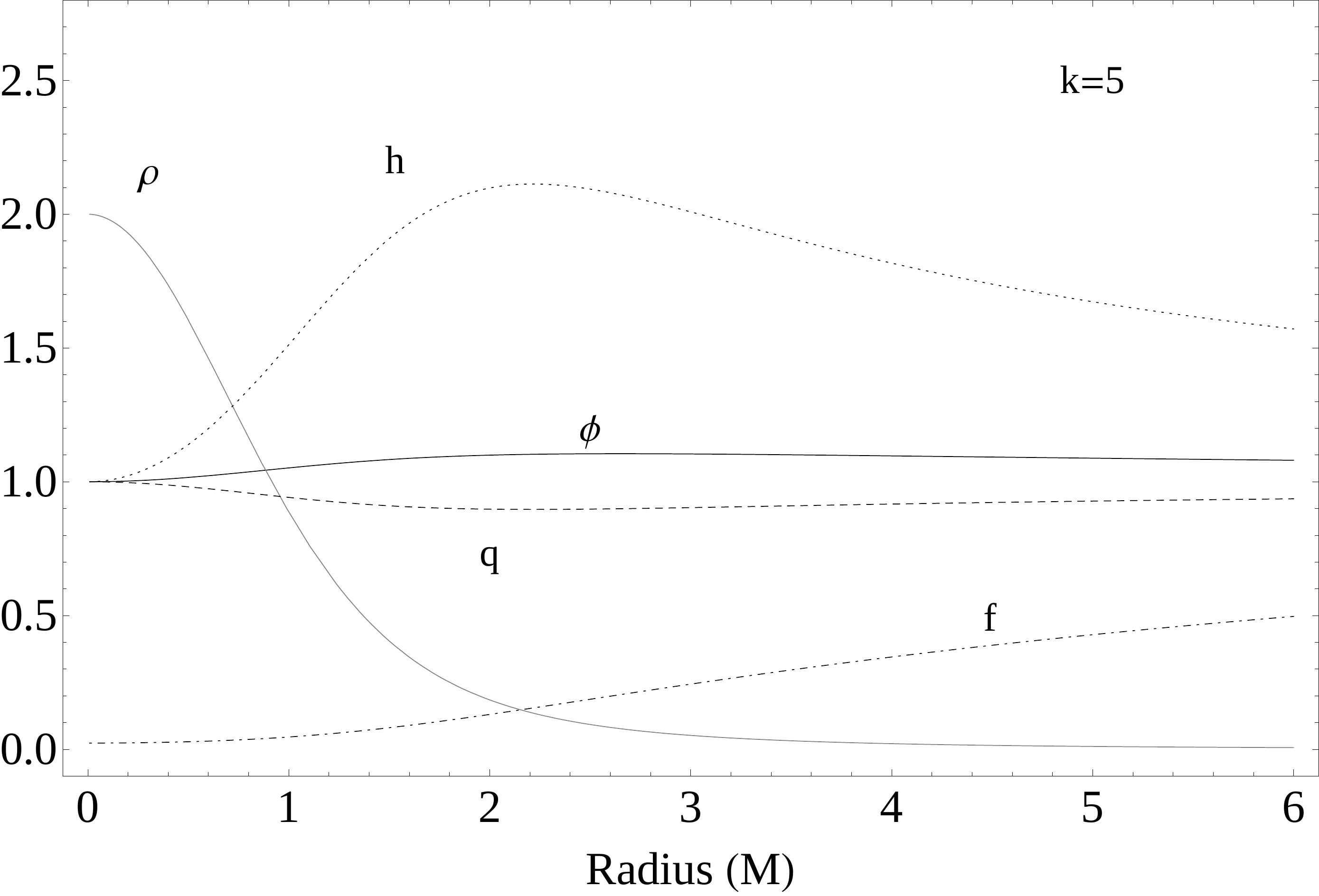}
\includegraphics[width=0.325\hsize,clip]{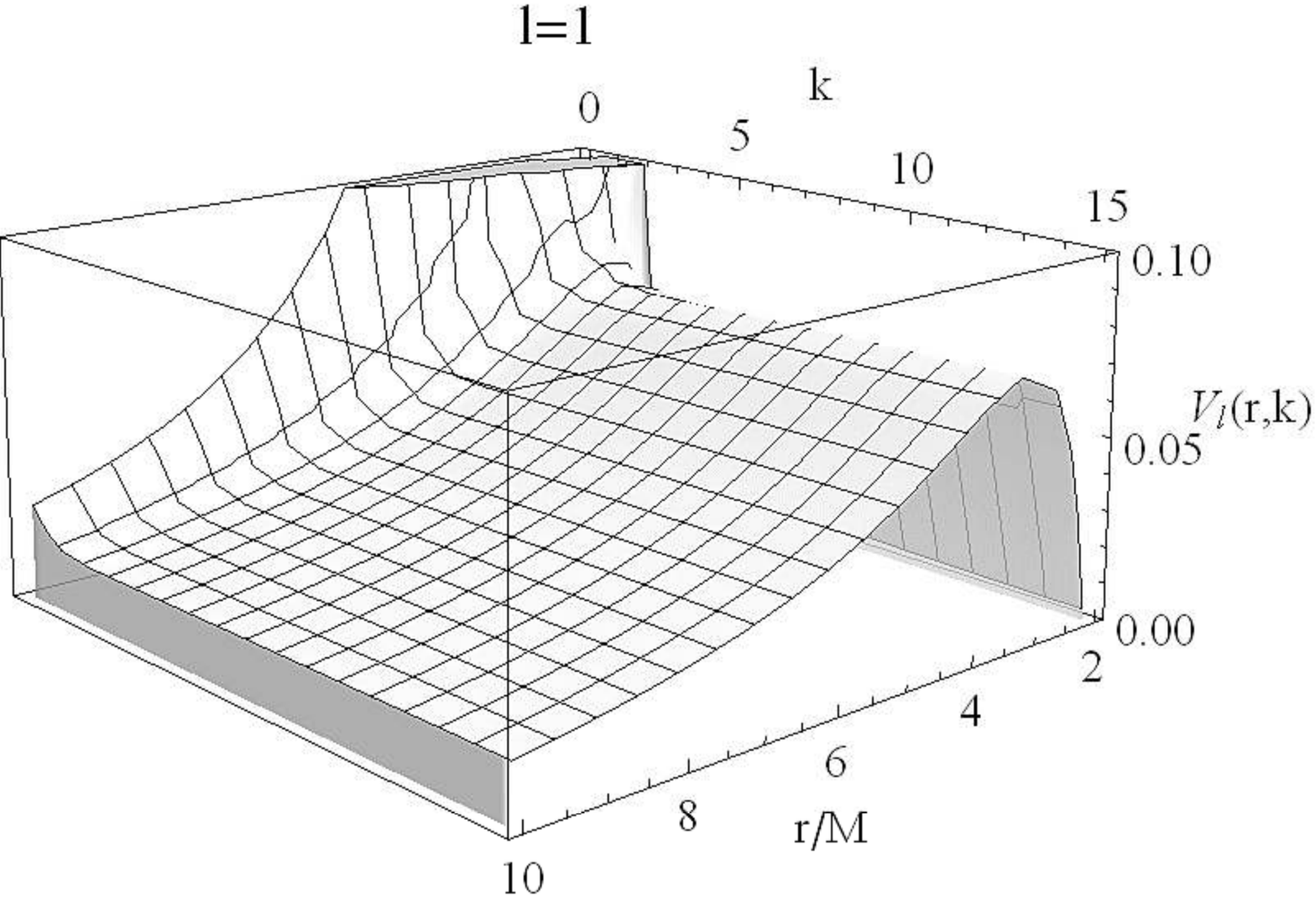}
\includegraphics[width=0.325\hsize,clip]{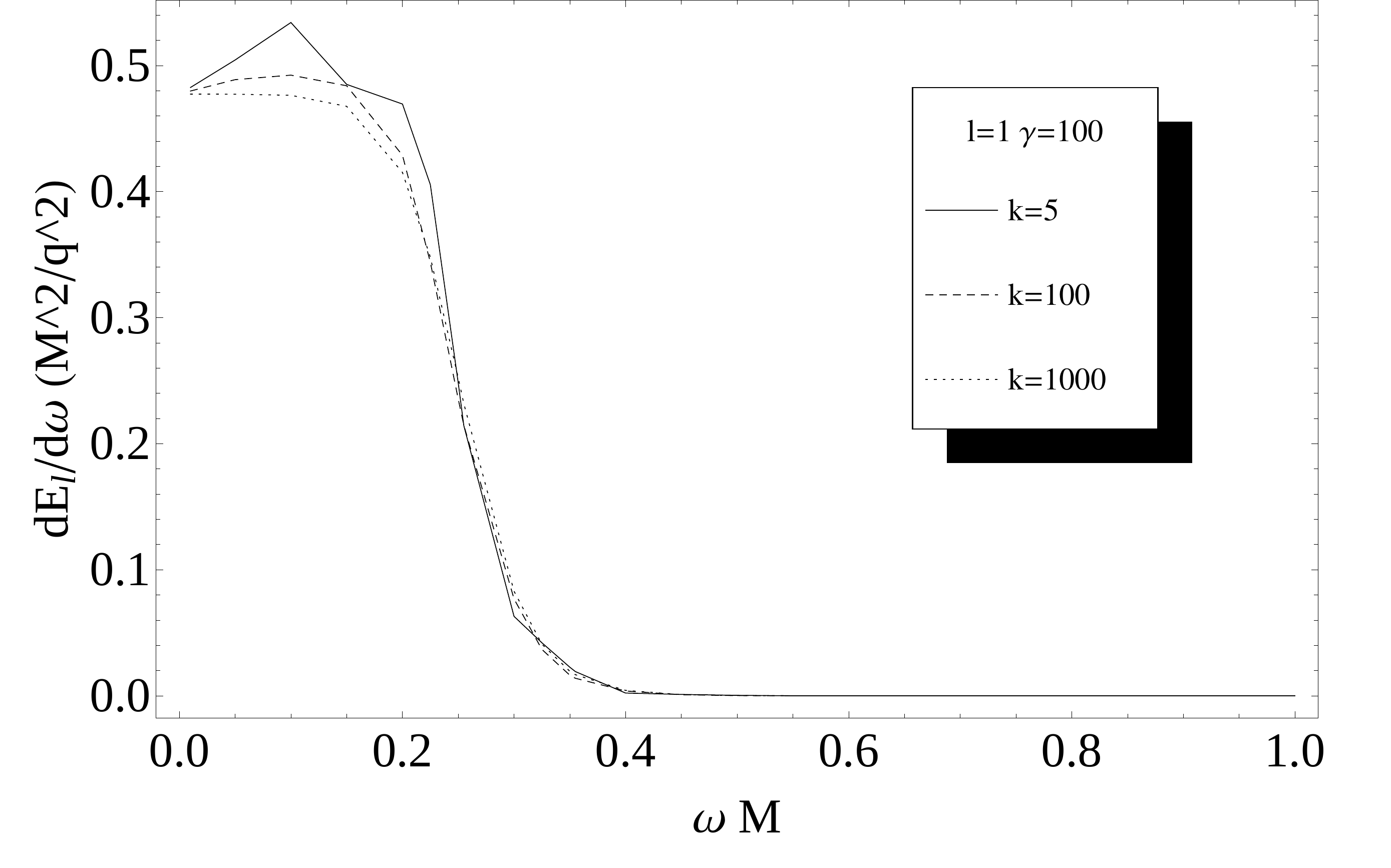}
\caption{Metric coefficients, $f$ (dotdashed line), $h$ (dotted line), $q$ (dashed line),  $\phi$ (black line) and the density $\rho$ (gray line), where $p=a\rho^b$, as functions of the radial coordinate, in unit of mass $M$, $b=1.27$.
$k=5$,  $a=0.45$  $\rho_0=2$ $x=9 \cdot 10^5$. Center: the potential $V_{l}(r,k)$ for   $l=1$ and as function of $k$ and $r/M$. Right: electromagnetic
energy spectra    for particle falling into a GSS background with different values of
$k$ is plotted for multipole $l=1$ and
$\gamma=100$.}\label{Fig:ste:R1}
\end{figure}

The KK equations  for  the  free electromagnetic case with $\vartheta=0$ are,

\begin{eqnarray}\label{kiastra1}
G_{\mu\nu}&=&\frac{1}{\phi}
\left(\nabla_{\mu}\partial_{\nu}\phi-g_{\mu\nu}\frac{8
\pi}{3}
\left(\rho-3p\right) +8\pi  T_{\mu\nu}\right),\\
\label{klastra2} \Box\phi&=&\frac{8 \pi}{3}
\left(\rho-3p\right), \quad
\nabla_{\rho}\left(T^{\mu \rho}\right)=0,
\end{eqnarray}
with  (4D) perfect fluid energy momentum tensor, $
T_{\mu\nu}=(p+\rho) u_{\mu} u_{\nu} -pg_{\mu\nu}
$, where  the pressure $p$ and the density  $\rho$ are related by the  equation of
state $p=a\rho^b$ with $(a,b)$ constants.
The
metric    has the  form
\begin{equation}\label{MetriIntKK}
ds_{\ti{(5)}}^{2}=f dt^{2}-h dr^{2}-q d\Omega^{2}-\phi^2 (dx_{\ti{5}})^2,
\end{equation}
where  $(f, h, q, \phi)$ are functions of  $r$ only.
Some  results of the numerical integration are shown   in Fig.\il(\ref{Fig:ste:R1})-left: metric components  and the matter density  $\rho$ are plotted as function of the radial coordinate $r$: fixing $k=5$, some initial conditions are given as the  central density $\rho_0=2$ and the scalar field at center  $\phi_0$ with $\phi_0'$, both  have been set as function of a free parameter $x$, fixed a priori in order  to match the GSS on the star boundary, see for details \cite{Pugliese:2011yh}.
This stellar model  represents    a spherically symmetric and stationary  star in the 4D Universe, whose matter can be described   as a perfect fluid with a given polytropic equation of state.
However, with respect to the correspondent solution of the Einstein equations  that meet at the board the Schwarzschild metric, this KK star is surrounded by a claud of extended scalar matter.
This is  clearly a toy model, and yet it is still only a first attempt to study this kind of objects, and it should be studied with greater and  deeper detail   especially paying attention to the behavior of the  ordinary matter by which it is supposed that the star is done.
Another important issue concerns the equilibrium and the phases of the stellar evolution that eventually can lead to a collapse with the formation of a singularity.
However we have assumed that the stellar surface is at some radius $ r = R $, it should be also noted that  this radius  can be, in general, very close to the Schwarzschild radius $ r = 2M $.
\section{Electromagnetic radiation emitted by a radially falling particle}\label{Sec:perturbation}
In \cite{Pugliese:2011um} it  has been provided the profile of the EM emission spectrum of  a charged particle radially falling on a  GSS toward a KK star:
it was assumed that the source of the gravitational field was a 5D stellar object as discussed in Sec.\il(\ref{Sec:KK-Stars}).
This assumption  has led to some specific hypothesis on the boundary conditions for the differential equations describing the spectrum:  having in mind that if the star model, solution of the 5D equations with a perfect fluid,  ``differs little'' from  the analogue  4D model, we assume  the presence of a possible extra dimension went sought in the  ``small deformations'' of the spectrum as expected in the EM emission from  a charged particle infalling in the Schwarzschild spacetime.  Here we concern on the simplest scenario  where $A=0$.  Topic is discussed in more details in
the cited works \cite{Lacquaniti:2009rh,Lacquaniti:2009wc}.
We consider the  charged particle dynamics     in a
region $r>R\equiv2(1+10^{-a})M$, with $a>0$ constant and we  compare the
dynamics in GSS with the dynamics  in the Schwarzschild geometry.

\begin{figure}
\includegraphics[width=0.45\hsize,clip]{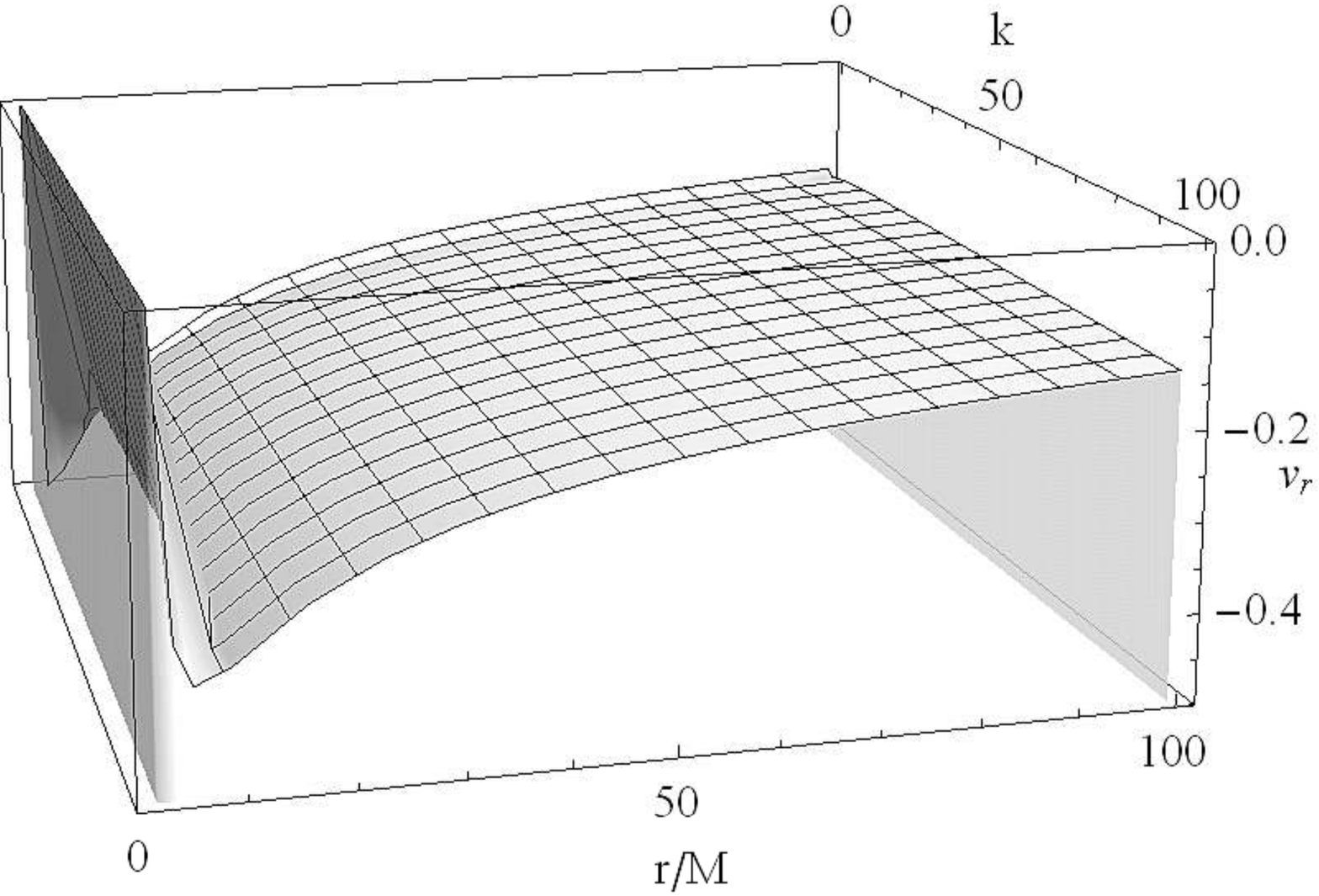}
\includegraphics[width=0.45\hsize,clip]{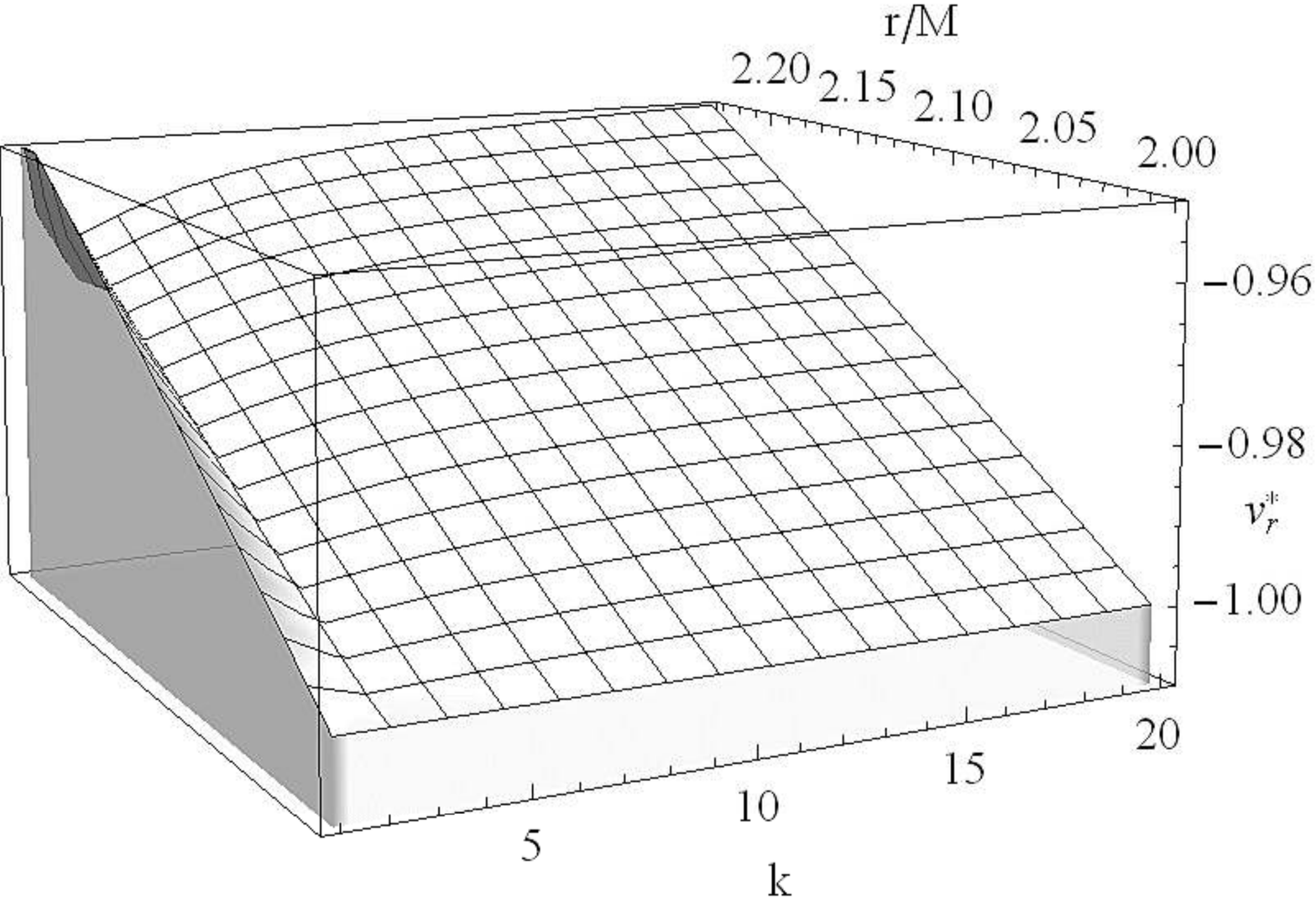}
\caption{Left: The picture shows the coordinate velocity component in the
$r$ direction $v_{r}\equiv r'=dr/dt$, for a particle starting from a
point at (spatial) infinity as function of $r/M$  and $k$. Right:  the locally measured radially velocity
$v_{r}^*\equiv r'_*=dr^*/dt$, where
$dr/dr^*=\sqrt{g_{tt}/g_{rr}}=\Delta^{\epsilon k-\epsilon/2}$, is plotted as function of  $r/M$ and $k$.}
\label{Figs:V_v_local}       
\end{figure}

In general, the motion of a radially falling particle is regulated by the
$t$ component and $r$ component of the geodesic equation
(\ref{Eq:PapA0}), with  $\dot{\theta}=\dot{\varphi}=0$, see for example \cite{Overduin:1998pn,Kalligas:1994vf}.
We introduce here the concept of the velocity  in the $r$ direction  $v_{r}\equiv r'=dr/dt$.
while the locally measured radial velocity is $v^*_{r}\equiv r'_*=dr^*/dt$
where $dr/dr^*=\sqrt{-g_{tt}/g_{rr}}=\Delta^{\epsilon k-\epsilon/2}$.
The radius  at which the radial velocity $v_{r}$ starts to
decreases will be denoted $\varrho$ \cite{Kalligas:1994vf}
and in the Schwarzschild limit it is $\varrho=6M$, otherwise
we have $\rho=\rho(k)<6M$.
The velocities $v_{r}$ and $v^*_{r}$ are functions  of $(r,k)$ and they tend to zero at
(spatial) infinity. The radial velocity $v_{r}$ goes to zero in the
limit $r\rightarrow2M$, and the local measured velocity $v^*_{r}$
goes to $(-1)$ for $r\rightarrow2M$, see Figs.\il(\ref{Figs:V_v_local}) and
\cite{Lacquaniti:2009yy,Lacquaniti:2009cr,Lacquaniti:2009rh,Lacquaniti:2009wc}.
The charged test particle motion and the emitted EM radiation is
considered as perturbations of the background metric, the EM radiation is regulated by
\begin{equation}\label{1801}
\partial_{\nu}\left(\sqrt{-g_4}\phi^3f^{\mu\nu}\right)= 4 \pi\sqrt{-g_4}
J^{\mu},
\end{equation}
where $f^{\mu\nu}$ is the Faraday tensor
  expanded in (4D) tensor
harmonics $(Y^*_{lm}(\Omega(t)))$, see
\cite{Lacquaniti:2009rq}. In agreement with the Zerilli
notation \cite{Zerilli:1971wd,Zerilli:1974ai,RuRR} $(\tilde{f}^{\mu\nu})$ denotes the radial
(angle independent) part     of the Faraday tensor, while the  $J^{\mu}$ is the  4D current \cite{Lacquaniti:2009rq,Lacquaniti:2009cr}, thus
according with the Zerilli's procedure, we find the following equation
\begin{equation}\label{Eq:q_u_a}
\frac{d^2\tilde{f}(\omega,r)}{dr_{*}^2}
+\left[\omega^2-V_{l}(r,k)\right]\tilde{f}(\omega,r)=S_{l}(\omega,r),
\end{equation}
and $S_{l}(\omega,r)$ is the
 source  term explicitly given in \cite{Pugliese:2011um}, this  is a function of $r$ and
$\gamma\equiv\frac{1}{\sqrt{1-v_{\infty}^2}}$, where $v_{\infty}$ is
the (radial)    particle velocity  at spatial infinity and of the
potential $V_{l}(r,k)$:
\begin{equation}
\label{Eq:vek}
V_{l}(r,k)=\frac{l(l+1)\Delta^{2\epsilon k-\epsilon-1}}
{r^2}+V_{0}(r,k),
\end{equation}
where
\begin{equation}\label{Eq:go_2010}
V_{0}(r,k)\equiv\frac{3\epsilon }{4}\frac{M^2\Delta^{2\epsilon
k-\epsilon-2}} {r^4}\left[4\left(\epsilon
k+1-\frac{r}{M}\right)+\epsilon\right],
\end{equation}
{(see\cite{Zerilli:1971wd,Zerilli:1974ai,RuRR,Pugliese:2011um} for the  current use of the notation)}.
In \cite{Pugliese:2011um} the Sturm Liouville problem  for the function
$\tilde{f}(\omega,r)$    was solved using  the Green's
functions method. First the  associated homogeneous ($S_{l}(\omega,r)=0$)
 equation has been  solved.  We used the solutions of this equation   to
build the Green's function $G(r,r')$ of the problem.
 It was  required that the outgoing radiation
is purely oscillating  at infinity.
The  energy spectra at (spatial) infinity was therefore given by
\begin{equation}\label{Eq:ini_fm}
\frac{dE}{d\omega}=\sum_{l}\frac{dE_{l}}{d\omega}=\sum_{l}\frac{l(l+1)}{2
\pi}\left|A^{out}_{l}(\omega)\right|^{2},
\end{equation}
where $\omega\geq0$ and  where $A^{out}_{l}$ is a function of
$\omega$ only.
Adopting a
Runge Kutta method, we first integrated the homogeneous equation for
a fixed value of $k$ to obtain an evaluation of the function $f_l$ with the
required boundary condition. The numerical
integration started at $R=(2+10^{-6})M$, that is a source boundary very close to the Schwarzschild limit $r=2M$. It is stopped at large
values of $r$ as well as the integrals converge.
We integrated  Eq.\il(\ref{Eq:q_u_a})  numerically
for $r>R$ and for $k={1000,10,5}$. The potential is plotted
in Fig.\il(\ref{Fig:ste:R1})-center.

 Once the values of the parameters $(\gamma,l)$  are
fixed, the integration has been performed for many fixed values of
$\omega$. It was  so proved that the spectra profile, at large $k$ (say $k\gg5$) substantially coincides with the one emitted by a charged particle falling   in the Schwarzschild background \cite{Cardoso:2003cn}.
 For a fixed $k$, we note here that
the homogeneous equation is a ``Schrodinger-like'' equation for the
function $\tilde{f}(\omega,r^*(r))$ with a potential
$V_{l}(r^*(r),k)$.
However, to fix the boundary conditions for a fixed $k$ is clearly important  the analysis of the potential $V_{l} (r, k) $ at infinity and at the starting point of integration. This function  is  plotted in Fig.\il(\ref{Fig:ste:R1})-center showing the  dependence from $ r/M $ and the parameter $ k $.
Asymptotically, for large value of $r$, the
potential $V_{l}(r,k)$ decreases to zero, it vanishes as $r$ approaches $2M$ in the Schwarzschild limit.
Far from this limit, ie for small values of $ k $ on $ r = R $, the  starting point of the integration, the  potential is finite and very small (see details in \cite{Pugliese:2011um}), this point in our model coincides with the  boundary of the stellar object. We assume a  regular  condition on the stellar surface with respect to the radiation emitted by taking the potential   to be zero a $r=R $.
The homogeneous equation is then solved imposing that the solution is ingoing
at $R$ and a combination of ``incident'' and ``transmitted'' waves
at spatial infinity.
 We have taken the star surface  in $r=R$ is  totally absorbent respect to the  EM radiation that invests it, which therefore can be represented by the  boundary condition on the surface with totally ingoing component.
But within the context that we are analyzing the behavior of EM radiation which invests  the gravitational source, has clearly an evolution determined by the gravitational field generated by the KK source. In  Eq.\il(\ref{Eq:q_u_a}) this information is encoded  particularly in the potential, which is modulated by the distortion of the spacetime metric.
In this regards, we note here some considerations concerning the $V_l $. The emission spectrum, being  in fact governed  by the function $V_l $,  is determined  by the boundary conditions that, according to  this, are fixed.
From Eq.\il(\ref{Eq:vek}) the potential  depends on several variables, namely the radius $r$,  the multipole  order $l$,  the source mass $ M $ and the parameter  $ k $, which precisely sets the background metric, but it does  not directly depend on the scalar field $ \phi $ and then from the extra dimension. The peculiar behavior of the potential  at the Schwarzschild radius,  differs greatly from the potential   $V_l $  in  the Schwarzschild spacetime diverging ar $r=2M$. This fact  is not  directly attributable to a possible influence of the extra dimension, but rather to the nature of the naked singularity located in $r=2M $, this speculation seems confirmed by the fact that the potential is stabilized (at a variation of $k$) in that point, being  regularized to zero on the limit for the Schwarzschild metric parameters even  in the full (not 4D reduced) 5D model, that is even when the scale  of the extra dimension is a constant but $x_5\neq{\rm{constant}}$.
On the other hand, as discussed in Sec.\il(\ref{Sec:GSS}), the  scalar field that regulates the size  of the fifth dimension diverges as one approaches the singularity. Other examples of similar studies which discuss the emission and  electromagnetic propagation
in the spacetime with naked singularity may be found in \cite{Dotti:2008ta,Kovacs:2010xm,DiCriscienzo:2010gh,deFelice74,Casadio:2003iv,Virbhadra:2007kw,Virbhadra:2002ju,Pat2011,Pat2010,Joshi:2012mk,Patil:2011uf}.

A spectra profile is plotted in Fig.~(\ref{Fig:ste:R1})-left for $\gamma=100$ and $l=1$, see also \cite{Pugliese:2011um}.
A general feature of the emitted spectra  for the
Schwarzschild's case is to grow up to a critical value
(corresponding to a critical frequency $\omega$ for a fixed
multipole) and then rapidly flow down to zero\cite{Cardoso:2003cn}.
We found that  for a fixed $\gamma$, the spectra
profile coincides with that of the  Schwarzschild's case as well as $k$
is sufficiently large ($k>5$). However there is an increase of the energy emitted
rate  per frequency $dE/d\omega$, at fixed value of $\omega$ and  fixed multipole $l$. There are significant discrepancies with the
Schwarzschild's case for $k\leq5$ where, for low frequencies, a peak in the spectra profile
appears  also at large $\gamma$.
\section{Discussions and future perspectives}\label{Sec:concldisc}
A
nontrivial question concerning  any extra dimensional geometrical theory, arises on the non observation of the
extra space in a Universe that is viewed  according to the 4D model of the  General Relativity.
The compactified 5D Kaluza Klein theory provides a natural response to this problem  assuming   a  closed extra space,
with a volume much smaller then the minimal observed scale
in the present day experiment of high energy physics, say
$\mathcal{O}(10^{-18}\rm{cm})$.
There is a great deal of attention in  notice any discrepancies in the   current observations with respect to the   predictions of the standard theoretical   setup, that could be possibly explained by an  extra dimensional theory. Any experimental observation
 could provide then a strong
constraint concerning the  validity of the extra dimensional hypothesis.
In particular one can search for   new  phenomena, unpredicted  by the current models, but explained in a multi dimensional framework.
One can search for the extra dimension evidences  in the microscopic world
\cite{Kong:2010mh,Bhattacherjee:2010vm,Datta:2010us,Franceschini:2011wr} or also in the astrophysical scenarios
\cite{Peter:2012rz,Einasto:2009zd,Li:2011sd,Copeland:2006wr,Frieman:2008sn,Sahni:2004ai,Kamionkowski:2007wv}
In both cases, the validity of a multi dimensional theory is bounded for providing a theoretical model able  to reproduce  the  compatibility of the extra dimension in a world that looks like a
four dimensional one, requiring essentially the reproducibility of the reduced Universe  and its physics in a suitable limit.
In this respect a problem, in particular,   plagues the  5D KK models  concerning  the particle charge to mass ratio  as obtained from the geodesic motion in the standard approach:
it is generally assumed the existence of a 5D particle generalizing its concept from  the 4D Universe   to a 5D world  and assuming that the particle dynamics, not subjected to any forces  in the  compactified model, could  be simply described  with a  5D geodesic   generalized from the equivalent 4D geodesic. Basically  this approach states that  the particle motion  projected along the fifth dimension, when it is  not subjected to forces in the  KK spacetime, can be also a 5D geodesic.
Recently, contesting the validity of the 5D geodesic assumption  with respect to the projection on the compactified dimension, has emerged the need to address the issue of the  test particle motion  within the extra dimension with closed topology in a different way,  which takes into account the  quasi Plankian  scale of this extra space. This problem  was approached by a Papapetrou multipole expansion  assuming a generic energy momentum tensor  centered on the 4D trajectory.
The  resulted dynamic equations,  are very promising and albeit able to describe correctly the 4D dynamics, they have proved to solve in particular the problem of the charge to mass ratio.

In this paper we  adopt this  multipole approach  to the matter dynamics in  the 5D spacetime of the KK model.
We  reformulated the  test particle motion and the KK field equations  finally arriving to a  model of a 5D stellar configuration (KK star).
We have presented a review of some astrophysical phenomenology   that can lead, through the direct comparison between the observations and the model predictions, to test the multi dimension hypothesis  viability.
Following the works\cite{Lacquaniti:2010zz,Lacquaniti:2009rh,Pugliese:2011um,Pugliese:2011yh,Lacquaniti:2009wc} we explored, by  an effective potential approach to the motion, the dynamic stability  around very compact objects; the  stability regions and the energies of the moving bodies were  found to be  significantly different from those expected in the 4D model resulting from the Einstein theory.
In  \cite{Pugliese:2011um} and in Sec.\il(\ref{Sec:perturbation})  of this review we have showed that   the spectral analysis from the sources of the high energy Universe
could in principle provide   a   test to constrain the
multidimensional theories appearing in  a departing  of the EM emission
spectra from that expected in
 general relativistic calculations  (see also \cite{Matsuno:2011ca,Bonnevier:2011km,Becar:2011fc,Inte,Cardoso:2003cn,lw,ww}).
The construction of a stellar object in this extra dimensional scenario has consistently proved the existence of a source of the gravitational field, that looks, in the asymptotical limit of Minkowski spacetime as and ordinary star  described by the Einstein equations.
We have assumed particular spacetime symmetries such that this KK star can be conveniently viewed in comparison with the source of the Schwarzschild spacetime.

In conclusion  the results,  are  convincing and promising under many respects.
 First,  the revision   of the dynamics  in the compactified scenarios provided by the multipole expansion   is  capable to solve different problems  of the standard approach to the KK models (in particular the charge to mass ratio puzzle for elementary particle).
This reinterpretation started from a reasonable discussion on the laws of motion in the spacetime geometries   with a Planckian scale dimension, investigating   the consequences in the dynamics as seen in the macroscopic world.

The propose to search the effects of a possible extra dimension in the  high energy Universe  is convincing
as confirmed by the results obtained  also in the  simple analysis of the test particle motion  and  the emission spectrum: the discrepancies  highlighted between the predictions of these models and the general relativistic ones  are small but may be detectable.

\section*{Acknowledgments}
This work has been developed in the framework of the CGW Collaboration
(www.cgwcollaboration.it). DP gratefully acknowledges financial support from the Angelo
Della Riccia Foundation and  wishes to thank the Blanceflor Boncompagni-Ludovisi, n\'ee Bildt 2012.


\begin{thebibliography}{99}

\bibitem{Bergamini:1984gx}
  R.~Bergamini and C.~A.~Orzalesi,
  Phys.\ Lett.\  B {\bf 135} (1984) 38.

\bibitem{Aranda:2009wh}
  A.~Aranda, J.~L.~Diaz-Cruz and A.~D.~Rojas,
  Phys.\ Rev.\  D {\bf 80} (2009) 085027.


\bibitem{Gunther:2002mm}
  U.~Gunther, P.~Moniz and A.~Zhuk,
  Astrophys.\ Space Sci.\  {\bf 283} (2003) 679.


\bibitem{Brax:2003fv}
  P.~Brax and C.~van de Bruck,
  Class.\ Quant.\ Grav.\  {\bf 20} (2003) R201.



\bibitem{Langlois:2002bb}
  D.~Langlois,
  Prog.\ Theor.\ Phys.\ Suppl.\  {\bf 148} (2003) 181.


\bibitem{Papantonopoulos:2002ew}
  E.~Papantonopoulos,
  Lect.\ Notes Phys.\  {\bf 592} (2002) 458.

\bibitem{Gun.Star.Zhuk04}
U. Günther, A. Starobinsky and A. Zhuk,
Phys. Rev. D 69, 044003 (2004).

\bibitem{Ar-Gi-Smo-Yo-2013}
M. Arai, Gi-Chol Cho, K. Smolek, and K. Yoneyama
Phys. Rev. D 87, 016010 (2013).

\bibitem{Sun:2012xt}
  H.~Sun, Y.~-J.~Zhou and H.~Chen,
  Eur.\ Phys.\ J.\  {\bf 72} (2012) 2011.

\bibitem{Choudhury:2011jk}
  D.~Choudhury, A.~Datta, D.~K.~Ghosh and K.~Ghosh,
  JHEP {\bf 1204} (2012) 057.


\bibitem{deBlas:2012qf}
  J.~de Blas, A.~Delgado, B.~Ostdiek and A.~de la Puente,
  Phys.\ Rev.\ D {\bf 86} (2012) 015028.




\bibitem{Zhou-Fei-Gan-You-Lei-2012}
Li Xiao-Zhou, Duan Peng-Fei, Ma Wen-Gan, Zhang Ren-You, and Guo Lei,
Phys. Rev. D 86, 095008 (2012).

\bibitem{Gerwick:2011jw}
  E.~Gerwick, D.~Litim and T.~Plehn,
  Phys.\ Rev.\ D {\bf 83} (2011) 084048.






\bibitem{Arrenberg:2008wy}
  S.~Arrenberg, L.~Baudis, K.~Kong, K.~T.~Matchev and J.~Yoo,
  Phys.\ Rev.\ D {\bf 78} (2008) 056002.





\bibitem{Calmet:2009yw}
  X.~Calmet, P.~de Aquino and T.~G.~Rizzo,
  Phys.\ Lett.\ B {\bf 682} (2010) 446.


\bibitem{Matsuno:2011ca}
  K.~Matsuno and K.~Umetsu,
  Phys.\ Rev.\  D {\bf 83} (2011) 064016.


\bibitem{Rizzo:1999qb}
  T.~G.~Rizzo,
  AIP Conf.\ Proc.\  {\bf 530} (2000) 290.
%

\bibitem{Eingorn:2010wi}
  M.~Eingorn and A.~Zhuk,
  Phys.\ Rev.\  D {\bf 83} (2011) 044005.



\bibitem{Moutsopoulos:2011ez}
  G.~Moutsopoulos and P.~Ritter,
  arXiv:1103.0152.


\bibitem{Stelea:2009ur}
  C.~Stelea, K.~Schleich and D.~Witt,
  Phys.\ Rev.\  D {\bf 83} (2011) 084037.






\bibitem{Bonnevier:2011km}
  J.~Bonnevier, H.~Melbeus, A.~Merle and T.~Ohlsson,
  arXiv:1104.1430.

\bibitem{Okawa:2011fv}
  H.~Okawa, K.~i.~Nakao and M.~Shibata,
  arXiv:1105.3331 [gr-qc].

\bibitem{Tomizawa:2011mc}
  S.~Tomizawa and H.~Ishihara,
  arXiv:1104.1468.

 \bibitem{Moon:2011sz}
  T.~Moon, Y.~S.~Myung and E.~J.~Son,
  arXiv:1104.1908.


\bibitem{Becar:2011fc}
  R.~Becar and P.~A.~Gonzalez,
  arXiv:1104.0356.

\bibitem{Yamada:2011br}
  Y.~Yamada and H.~a.~Shinkai,
  Phys.\ Rev.\  D {\bf 83} (2011) 064006.

\bibitem{Inte}
Q. Jiang, S.Yang, H. Li, Int. Journ. of Theoretical Physics, Vol. 45, No. 9, (2006).







\bibitem{Lacquaniti:2010zz}
  V.~Lacquaniti, G.~Montani and D.~Pugliese,
  Gen.\ Rel.\ Grav.\  {\bf 43}, 4, (2011) 1103.


\bibitem{Lacquaniti:2009rh}
  V.~Lacquaniti, G.~Montani, D.~Pugliese,
  arXiv:0911.4168 [gr-qc].


\bibitem{Pugliese:2011um}
  D.~Pugliese, G.~Montani and V.~Lacquaniti,
  Eur.\ Phys.\ J.\ C {\bf 71} (2011) 1747.




\bibitem{Pugliese:2011yh}
  D.~Pugliese and G.~Montani,
  Eur.\ Phys.\ J.\ C {\bf 71} (2011) 1638.


\bibitem{Lacquaniti:2009wc}
  V.~Lacquaniti, G.~Montani, D.~Pugliese and R.~Ruffini,
  arXiv:0912.2408 [gr-qc].








\bibitem{PaulBC}	
B. C. Paul,
Int.\ J.\ Mod.\ Phys.\  D, Vol. 13, Issue 02, pp. 229-238 (2004).



\bibitem{Hannestad:2003yd}
  S.~Hannestad and G.~G.~Raffelt,
  Phys.\ Rev.\  D {\bf 67} (2003) 125008
  [Erratum-ibid.\  D {\bf 69} (2004) 029901].

%
\bibitem{Patel:2001jw}
  L.~K.~Patel and G.~P.~Singh,
  Grav.\ Cosmol.\  {\bf 7}, 52 (2001).

\bibitem{cha}
P. K. Chattopadhyay,  B. C. Paul,
PRAMANA, journal physics, Vol. 74, No. 4, 2010
pp. 513-523.


\bibitem{K1}
T.~Kaluza, \emph{On the Unity Problem of Physics}, Sitzungseber.\ Press.\
Akad.\ Wiss.\ Phys.\
Math. ,1921.

\bibitem{K2}
O.~Klein, Z.F.Physik, 37, 1926.

\bibitem{K3}
O.~Klein, Nature, 118, 1926.


\bibitem{Lacquaniti:2009cr}
  V.~Lacquaniti and G.~Montani,
  arXiv:0906.2231 [gr-qc].




\bibitem{Lacquaniti:2009rq}
  V.~Lacquaniti and G.~Montani,
  Mod.\ Phys.\ Lett.\  A {\bf 24} (2009) 1565.

\bibitem{Lacquaniti:2009yy}
  V.~Lacquaniti and G.~Montani,
  Int.\ J.\ Mod.\ Phys.\  D {\bf 18} (2009) 929.






\bibitem{Cianfrani:2007ep}
  F.~Cianfrani, I.~Milillo and G.~Montani,
  Phys.\ Lett.\ A {\bf 366} (2007) 7.



%

  \bibitem{Overduin:1998pn}
  J.~M.~Overduin and P.~S.~Wesson,
  Phys.\ Rept.\  {\bf 283} (1997) 303.


%
\bibitem{Wesson:1999nq}
  P.~S.~Wesson,
 ``Space - time - matter: Modern Kaluza-Klein theory,''
{\it  Singapore, Singapore: World Scientific (1999) 209 p}.



\bibitem{Bailin:1987jd}
  D.~Bailin and A.~Love,
  Rept.\ Prog.\ Phys.\  {\bf 50} (1987) 1087.

 \bibitem{Lb-gr-a}\emph{Modern Kaluza-Klein Theories}, edited by T. Applequist, A. Chodos,
and P.G.O. Freund, Addison-Welsey, Menlo Park, 1987.

  \bibitem{papapetrou}
A. Papapetrou, \emph{Proc. Phys. Soc.} \textbf{64}, 57 (1951).


\bibitem{Gross:1983hb}
  D.~J.~Gross and M.~J.~Perry,
  Nucl.\ Phys.\  B {\bf 226} (1983) 29.

\bibitem{sorkin}
R. D. Sorkin,  \emph{Phys. Rev. Lett.}, \textbf{ 51}, (1983) 87.

\bibitem{davisonowen}
A. Davidson, D. A. Owen, \emph{Phys. Lett.}, \textbf{155B}, (1985)
247.

\bibitem{Xu:2007dc}
  P.~Xu and Y.~g.~Ma,
  Phys.\ Lett.\  B {\bf 656} (2007) 165.





\bibitem{PoncedeLeon:2006xs}
  J.~Ponce de Leon,
  Int.\ J.\ Mod.\ Phys.\  D {\bf 17} (2008) 237.

\bibitem{PoncedeLeon:2007bm}
  J.~Ponce de Leon,
  Int.\ J.\ Mod.\ Phys.\  D {\bf 18} (2009) 251.


\bibitem{Overduin:2000gr}
  J.~M.~Overduin,
  Phys.\ Rev.\  D {\bf 62} (2000) 102001.


\bibitem{MisThoWhe73}
C.~W. Misner, K.~S. Thorne, \& J.~A. Wheeler,
\newblock {\em Gravitation},
\newblock W. H. Freeman, 1973.




\bibitem{Kalligas:1994vf}
  D.~Kalligas, P.~S.~Wesson and C.~W.~F.~Everitt,
  Astrophys.\ J.\  {\bf 439} (1994) 548.




\bibitem{Zerilli:1971wd}
  F.~J.~Zerilli,
  Phys.\ Rev.\   D {\bf 2} (1970) 2141.

\bibitem{Zerilli:1974ai}
  F.~J.~Zerilli,
Phys.\ Rev.\   D {\bf 9} (1974) 860.

 \bibitem{RuRR}
  \emph{Les astres occlus-Les Houches}, C. and B.S. De Witt (eds.), Gordon and Breach (New York, 1973).

\bibitem{Cardoso:2003cn}
  V.~Cardoso, J.~P.~S.~Lemos and S.~Yoshida,
  Phys.\ Rev.\   D {\bf 68} (2003) 084011.





\bibitem{Dotti:2008ta}
  G.~Dotti and R.~J.~Gleiser,
  Class.\ Quant.\ Grav.\  {\bf 26} (2009) 215002.

\bibitem{Kovacs:2010xm}
  Z.~Kovacs and T.~Harko,
  Phys.\ Rev.\ D {\bf 82} (2010) 124047.



\bibitem{DiCriscienzo:2010gh}
  R.~Di Criscienzo, L.~Vanzo and S.~Zerbini,
  JHEP {\bf 1005} (2010) 092.




  \bibitem{deFelice74}
F. de Felice,   	
Astronomy and Astrophysics, Vol. 34, p. 15 (1974).

\bibitem{Casadio:2003iv}
  R.~Casadio, S.~Fabi and B.~Harms,
  Phys.\ Rev.\ D {\bf 70} (2004) 044026.


\bibitem{Virbhadra:2007kw}
  K.~S.~Virbhadra and C.~R.~Keeton,
  Phys.\ Rev.\  D {\bf 77} (2008) 124014.



\bibitem{Virbhadra:2002ju}
  K.~S.~Virbhadra and G.~F.~R.~Ellis,
  Phys.\ Rev.\  D {\bf 65} (2002) 103004.

\bibitem{Pat2011}
M. Patil, P.S. Joshi, D. Malafarina,
Phys.Rev. D83 064007 (2011).




  \bibitem{Pat2010} M. Patil, P. S. Joshi,
Phys.Rev. D82 104049 (2010).

\bibitem{Joshi:2012mk}
  P.~S.~Joshi and D.~Malafarina,
  Int.\ J.\ Mod.\ Phys.\ D {\bf 20} (2011) 2641.


\bibitem{Patil:2011uf}
  M.~Patil, P.~S.~Joshi, M.~Kimura and K.~-i.~Nakao,
  Phys.\ Rev.\ D {\bf 86} (2012) 084023.



\bibitem{Kong:2010mh}
  K.~Kong, K.~Matchev and G.~Servant,
  arXiv:1001.4801 [hep-ph].

\bibitem{Bhattacherjee:2010vm}
  B.~Bhattacherjee and K.~Ghosh,
with
  Phys.\ Rev.\  D {\bf 83} (2011) 034003.

\bibitem{Datta:2010us}
  A.~Datta, K.~Kong and K.~T.~Matchev,
  New J.\ Phys.\  {\bf 12} (2010) 075017.


\bibitem{Franceschini:2011wr}
  R.~Franceschini, G.~F.~Giudice, P.~P.~Giardino, P.~Lodone and
A.~Strumia,
  arXiv:1101.4919 [hep-ph].




\bibitem{Peter:2012rz}
  A.~H.~G.~Peter,
  arXiv:1201.3942 [astro-ph.CO].


\bibitem{Einasto:2009zd}
  J.~Einasto,
  arXiv:0901.0632 [astro-ph.CO].

\bibitem{Li:2011sd}
  M.~Li, X.~-D.~Li, S.~Wang and Y.~Wang,
  Commun.\ Theor.\ Phys.\  {\bf 56} (2011) 525.


\bibitem{Copeland:2006wr}
  E.~J.~Copeland, M.~Sami and S.~Tsujikawa,
  Int.\ J.\ Mod.\ Phys.\ D {\bf 15} (2006) 1753.


\bibitem{Frieman:2008sn}
  J.~Frieman, M.~Turner and D.~Huterer,
  Ann.\ Rev.\ Astron.\ Astrophys.\  {\bf 46} (2008) 385.


\bibitem{Sahni:2004ai}
  V.~Sahni,
  Lect.\ Notes Phys.\  {\bf 653} (2004) 141.

\bibitem{Kamionkowski:2007wv}
  M.~Kamionkowski,
  arXiv:0706.2986 [astro-ph].

\bibitem{lw}
H. Liu and P. S. Wesson,
J. Math. Phys. 42, 4963 (2001).





\bibitem{ww}
M. Cassé, J. Paul, G. Bertone,  G. Sigl,
Phys. Rev. Lett. 2004 Mar 19;92(11):111102.



 \end{thebibliography}
\end{document}